\documentclass[pdflatex,oneside,sn-nature]{sn-jnl}

\usepackage{graphicx}%
\usepackage{multirow}%
\usepackage{amsmath,amssymb,amsfonts}%
\usepackage{amsthm}%
\usepackage{mathrsfs}%
\usepackage[title]{appendix}%
\usepackage{xcolor}%
\usepackage{textcomp}%
\usepackage{manyfoot}%
\usepackage{booktabs}%
\usepackage{algorithm}%
\usepackage{algorithmicx}%
\usepackage{algpseudocode}%
\usepackage{listings}%
\usepackage{fix-cm}%
\usepackage{caption}
\usepackage{hyperref}

\begin{document}

\title[Article Title]{Planets Across Space and Time (PAST). VI. \\Age Dependence of the Occurrence and Architecture of Ultra-Short-Period Planet Systems}

\author[1,2]{\fnm{Pei-Wei} \sur{Tu}}\email{pwtu@smail.nju.edu.cn}

\author*[1,2]{\fnm{Ji-Wei} \sur{Xie}}\email{jwxie@nju.edu.cn}

\author[1,2]{\fnm{Di-Chang} \sur{Chen}}\email{dcchen@nju.edu.cn}

\author[1,2]{\fnm{Ji-Lin} \sur{Zhou}}\email{zhoujl@nju.edu.cn}

\affil*[1]{\orgdiv{School of Astronomy and Space Science}, \orgname{Nanjing University}, \orgaddress{\postcode{210023}, \state{Nanjing}, \country{China}}}

\affil[2]{\orgdiv{Key Laboratory of Modern Astronomy and Astrophysics}, \orgname{Ministry of Education}, \orgaddress{\postcode{210023}, \state{Nanjing}, \country{China}}}

\abstract{Ultra-short-period (USP) planets, with orbital periods shorter than one day, represent a unique class of exoplanets whose origin remains puzzling. 
Determining their age distribution and temporal evolution is vital for uncovering their formation and evolutionary pathways. 
Using a sample of over 1,000 short-period planets around Sun-like stars, we find that the host stars of USP planets are relatively older and have a higher prevalence in the Galactic thick disk compared to stars hosting other short-period planets. 
Furthermore, we find that the occurrence of USP planets increases with stellar age and uncover evidence indicating that USP planetary system architectures evolve on Gyr timescales. 
This includes a distinct dip-pileup in period distributions around $\sim$1 day and an expansion of orbital spacings with time. 
In addition, younger USP planet systems are observed to have fewer multiple transiting planets, implying fewer nearby companions and/or larger mutual orbital inclinations. 
Our findings suggest that USP planets continuously form through inward migration driven by tidal dissipation over Gyr timescales, and that younger and older USP planets may have originated via different specific tidal migration pathways.
}


\maketitle

\clearpage

{\bf \large \center Table of Contents\\}
\noindent
{\bf \large Main Text} \\
\hyperlink{Sample_selection}{Sample selection} \hfill \pageref{sec:Sample_selection}\\
\hyperlink{Analyses_and_results}{Analyses and results} \hfill \pageref{sec:Analyses_and_results}
 
   \hyperlink{R1}{USP planet hosts are dynamically hotter and older (Fig. 1-2)} \hfill \pageref{sec:R1}
   
   \hyperlink{R2}{USP planet frequency increases with time (Fig. 3)} \hfill \pageref{sec:R2}
   
   \hyperlink{R3}{USP planetary system architecture changes with time (Fig. 4)} \hfill \pageref{sec:R3}\\
\hyperlink{Discussions_and_conclusions}{Discussions and conclusions} \hfill \pageref{sec:Discussions_and_conclusions}\\
\\
{\bf \large Methods \& Supplementary Material} \\
\hyperlink{Data_sample}{{\bf 1 Data sample}} \hfill \pageref{sec:Data_sample} \\
\hyperlink{1.1}{\hspace{2em}1.1 Sample selection (Supplementary Table 1)} \hfill \pageref{sec:1.1}

\hyperlink{1.1.1}{\hspace{2em}1.1.1 Obtaining the hot planet host sample} \hfill \pageref{sec:1.1.1}

\hyperlink{1.1.2}{\hspace{2em}1.1.2 Obtaining astrometric parameters} \hfill \pageref{sec:1.1.2}

\hyperlink{1.1.3}{\hspace{2em}1.1.3 Obtaining spectroscopic parameters} \hfill \pageref{sec:1.1.3}\\
\hyperlink{1.2}{\hspace{2em}1.2 Calibrating parameters from various sources (Supplementary Fig. 1)} \hfill \pageref{sec:1.2}\\
\hyperlink{1.3}{\hspace{2em}1.3 Classifying galactic composition} \hfill \pageref{sec:1.3}\\
\hyperlink{1.4}{\hspace{2em}1.4 Final samples (Supplementary Fig. 2-3)} \hfill \pageref{sec:1.4}\\
\hyperlink{Stellar_parameter_control}{{\bf 2 Stellar parameter control} (Supplementary Fig. 4)} \hfill \pageref{sec:Stellar_parameter_control} \\
\hyperlink{Metallicity}{{\bf 3 Metallicity distribution of hot planet hosts} (Supplementary Fig. 5)} \hfill \pageref{sec:Metallicity} \\
\hyperlink{Kinematic}{{\bf 4 Kinematic properties of hot planet hosts}} \hfill \pageref{sec:Kinematic} \\
\hyperlink{4.1}{\hspace{2em}4.1 Comparison of total velocity and TD/D (Supplementary Fig. 6-7)} \hfill \pageref{sec:4.1}\\
\hyperlink{4.2}{\hspace{2em}4.2 Comparison of velocity dispersion distribution (Supplementary Fig. 8)} \hfill \pageref{sec:4.2}\\
\hyperlink{4.3}{\hspace{2em}4.3 Comparison of kinematic age distribution \\ \hspace*{2em}(Supplementary Fig. 9 and Supplementary Table 2)} \hfill \pageref{sec:4.3}\\
\hyperlink{Frequency}{{\bf 5 Frequency of USP planets as a function of age} (Supplementary Fig. 10-12)} \hfill \pageref{sec:Frequency} \\
\hyperlink{architecture}{{\bf 6 Age dependence of the architecture of USP planet systems} \\ (Supplementary Fig. 13-18)} \hfill \pageref{sec:architecture} \\
\hyperlink{6.1}{\hspace{2em}6.1 Orbital period distribution} \hfill \pageref{sec:6.1}\\
\hyperlink{6.2}{\hspace{2em}6.2 Orbital period ratio distribution} \hfill \pageref{sec:6.2}\\
\hyperlink{6.3}{\hspace{2em}6.3 Transiting planet multiplicity} \hfill \pageref{sec:6.3}\\
\hyperlink{Supplementary_discussion}{{\bf 7 Supplementary discussion}} \hfill \pageref{sec:Supplementary_discussion} \\
\hyperlink{7.1}{\hspace{2em}7.1 Validating the effects of detection completeness \\ \hspace*{2em}(Supplementary Fig. 19-21)} \hfill \pageref{sec:7.1}

\hyperlink{7.1.1}{\hspace{2em}7.1.1 The frequency of Kepler USP planets as a function of age} \hfill \pageref{sec:7.1.1}

\hyperlink{7.1.2}{\hspace{2em}7.1.2 Kepler USP planetary system architecture changes with time} \hfill \pageref{sec:7.1.2}

\hyperlink{7.1.3}{\hspace{2em}7.1.3 Section summary} \hfill \pageref{sec:7.1.3}

\clearpage

Ultra-short-period (USP) planets are typically defined as small planets with orbital periods $P\lesssim$ 1 day around their host stars. 
The existence of such planets, orbiting extremely close to their host stars, challenges the traditional understanding of planetary formation theories. 
From the discovery of the first USP planet \citep[CoRoT-7b,][]{2009A&A...506..287L} to the identification of over a hundred USP planets in {\it Kepler} data \citep{2014ApJ...787...47S,2021PSJ.....2..152A}, research on USP planets has garnered considerable attention in recent years, revealing some intriguing characteristics. 
For example, Winn et al. (2017) \citep{2017AJ....154...60W} shows that USP planets are not associated with metal-rich stars, distinguishing them as a distinct group from hot Jupiters in terms of their host stars' metallicity.
Most USP planets have compositions similar to Earth \citep{2019ApJ...883...79D}. They are more likely to be the remnants of photoevaporation from sub-Neptunes with H/He envelopes \citep{2017MNRAS.472..245L}, rather than the residual rocky cores of giant planets \citep{2017ApJ...846L..13K}.
Furthermore, Dai et al. (2018) \citep{2018ApJ...864L..38D} found that USP planets tend to have higher mutual inclinations ($\sim 5-15^{\circ}$) compared to most {\it Kepler} systems ($\lesssim 5^{\circ}$) and our solar system ($\sim 3^{\circ}$). 
Despite the growing research efforts aimed at uncovering the unique features of USP planets, their origins remain enigmatic \citep[see the review by][]{2018NewAR..83...37W}. 

A crucial aspect in elucidating the formation and evolutionary history of USP planets lies in their dependence on the age of their host stars. Exploring whether there exist disparities in the age distribution of host stars between USP planets and other short-period planets, such as hot Jupiters and hot small planets (e.g., hot super-Earths), offers valuable insights. 
Nevertheless, accurately estimating the ages of main-sequence host stars is a major challenge in studying the temporal evolution of USP planets. 
The commonly used isochrone fitting method can estimate the ages of a large number of individual main-sequence host stars, but typically with a large uncertainty of over $\sim50\%$ \citep[e.g.,][]{2020AJ....159..280B}. 

As an alternative, the average age of a group of stars can be statistically derived from their kinematics by employing the Age-Velocity dispersion Relation (AVR) \citep[e.g.,][]{1977A&A....60..263W,2009A&A...501..941H}.
Previously, in a series of papers on the Planet Across Space and Time project (hereafter referred to as PAST), we have refined the AVR to derive the kinematic ages with internal uncertainties of $\sim 10\%-20\%$ \citep{2021ApJ...909..115C}, and applied the kinematic method to delve into the age-dependency of super-Earths/sub-Neptunes \citep{2022AJ....163..249C,2023AJ....166..243Y} as well as hot Jupiters \citep{2023PNAS..12004179C}.
In this work, we extend the PAST series by statistically investigating the age distribution and temporal evolution of USP planets. 
Specifically, we seek to compare the kinematic ages of stars hosting USP planets with those hosting other hot planets, ultimately deriving the relative frequency and architecture of USP planetary systems as a function of stellar age.

\hypertarget{Sample_selection}{\section*{Sample selection}}
\phantomsection\label{sec:Sample_selection}

We initialize our sample of planets discovered by transit method and their host stars from the catalogs of confirmed planets and the {\it Kepler} DR 25 candidates on NASA exoplanet archive \citep[https://exoplanetarchive.ipac.caltech.edu;][]{2013PASP..125..989A}.
We exclude planets flagged in binary star systems and select only hot planets ($P < 10$ days) around Sun-like stars as our sample of planets. 
We crossmatch with Gaia DR3 to obtain astrometric parameters for the host stars and further crossmatch with LAMOST and CKS to obtain spectroscopic parameters. 
After calibrating the stellar spectroscopic parameters from different sources (\hyperlink{figure_s1}{Supplementary Fig. 1}), we then use methods from the PAST project \citep{2021ApJ...909..115C,2021AJ....162..100C} to obtain the kinematic properties of the stars, e.g., Galactic velocity and component membership (see \hyperlink{Data_sample}{Methods \S1} for more details on the sample selection as well as the characterization and calibration of planet host stars). 
Finally, our sample includes 826 stars in the Galactic disk hosting 975 hot planets. 
We classify planets into three categories: (1) USP planets with orbital periods shorter than 1 day and radii smaller than 6 Earth radii; (2) hot Jupiters (HJs) with orbital periods shorter than 10 days and radii greater than 6 Earth radii; and (3) hot small (HS) planets with orbital periods in the range of 1-10 days and radii smaller than 6 Earth radii.
Among these planets, there are 64 USP planets, 225 HJs, and 686 HS planets (see \hyperlink{table_s1}{Supplementary Table 1} and \hyperlink{figure_s2}{Supplementary Fig. 2}-\hyperlink{figure_s3}{3}). 

\hypertarget{Analyses_and_results}{\section*{Analyses and results}}
\phantomsection\label{sec:Analyses_and_results}

Utilizing Keck spectroscopy data, the previous study \citep{2017AJ....154...60W} revealed that USP planet host stars exhibit a metallicity distribution that differs significantly from that of HJ host stars, yet it closely aligns with the distribution observed in HS planet hosts. 
Here, we initially revisit this metallicity trend in our sample based on LAMOST spectroscopy data. 
 
\hyperlink{figure_s3}{Supplementary Fig. 3} depicts the distribution of effective temperatures ($T_{\rm eff}$) and surface gravitational accelerations ($\log g$) for the three categories of planetary host stars.
It is evident that their host stars exhibit significant differences in the $T_{\rm eff}$-$\log g$ plane; particularly, USP planet hosts tend to have higher $\log g$ but lower $T_{\rm eff}$. 
To mitigate the potential impact of these disparities on our subsequent analyses and results, we implement a stellar parameter control process (see \hyperlink{Stellar_parameter_control}{Methods \S2}) for the host stars of HJs and HS planets.
After the parameter control, these stars exhibit comparable distributions in the $T_{\rm eff}$-$\log g$ plane as compared to the USP planet hosts (\hyperlink{figure_s4}{Supplementary Fig. 4}). 

We then compare the metallicity ($\rm [Fe/H]$) distributions among the three planetary populations' host stars, as shown in \hyperlink{figure_s5}{Supplementary Fig. 5}. 
Notably, the $\rm [Fe/H]$ distribution of USP planet hosts differs from that of hot Jupiter hosts. 
The two-sample Kolmogorov–Smirnov (K-S) test conducted on these subsamples after (before) the stellar parameter control gives a $p-$value of 0.0590 (0.0611), corresponding to a difference of about $2 \sigma$.
Conversely, the $\rm [Fe/H]$ distribution of USP planet hosts is similar to that of hot small planet hosts, with a K-S test $p-$value of 0.2613 (0.1881) after (before) the parameter control. 

The above results align well with previous research conducted by Winn et al. (2017) \citep{2017AJ....154...60W}, further validating that the progenitors of USP planets are more likely HS planets, rather than HJs, from the perspective of host metallicity.
In the subsequent sections of this paper, we delve into the exploration of hot planets and their hosts from the perspective of age, which offers a direct insight into the emergence and evolution of USP planets.

\hypertarget{R1}{\subsection*{USP planet hosts are dynamically hotter and older.}}
\phantomsection\label{sec:R1}

\begin{figure*}[!t]
\centering
\includegraphics[width=\textwidth]{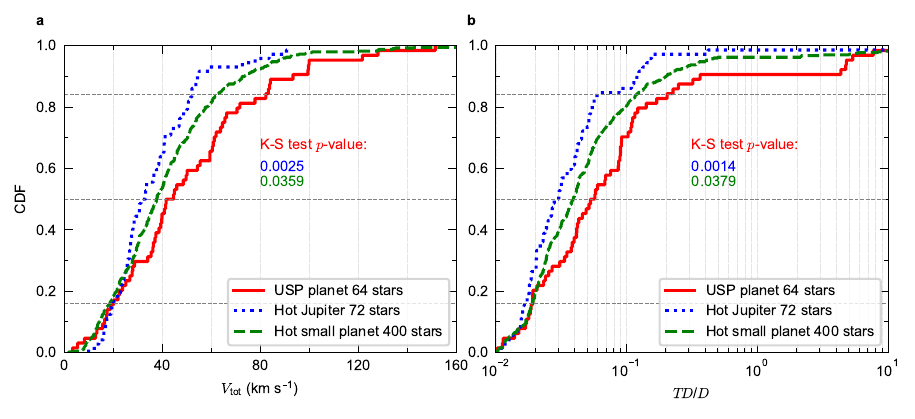}
\hypertarget{figure_1}{\caption{\textbf{Comparison of kinematic properties of hot planet hosts.}
Panels show the cumulative distributions of the total velocities ($V_{\rm tot}$, \textbf{a}) and the relative probabilities between thick disk to thin disk ($TD/D$, \textbf{b}) for the USP planet hosts (red solid lines), hot Jupiter hosts (blue dotted lines), and hot small planet hosts (green dashed lines). 
The horizontal gray dashed lines represent the 50$\pm$34.1 percentiles in the distribution. 
Each panel shows the K-S test $p-$values comparing hot Jupiter and hot small planet hosts to USP planet hosts. 
USP planet hosts exhibit higher values of $V_{\rm tot}$ and $TD/D$.
}}
\end{figure*}

\begin{figure*}[!t]
\centering
\includegraphics[width=\textwidth]{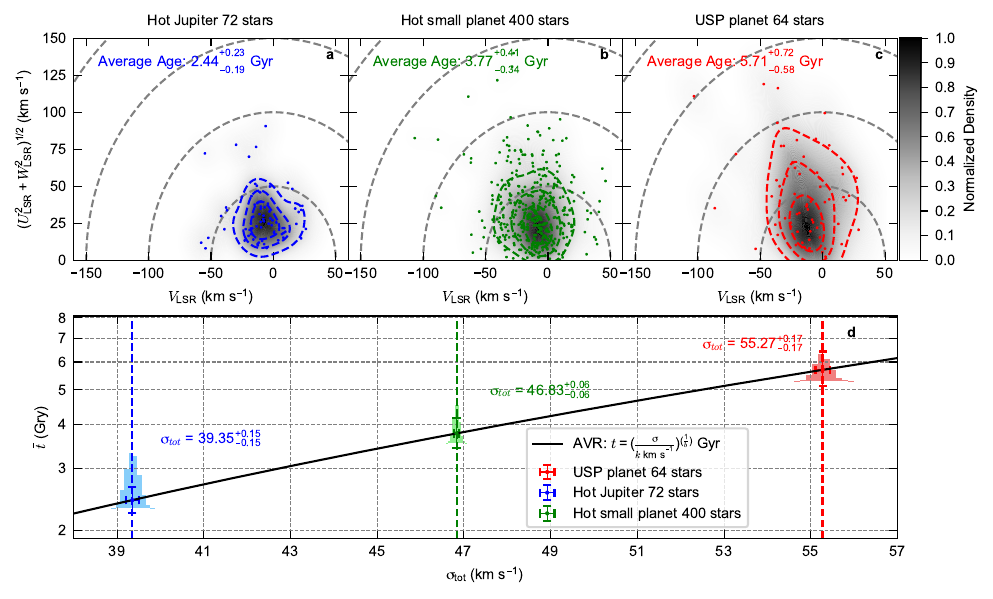}
\hypertarget{figure_2}{\caption{\textbf{Comparison of the Galactic velocity distributions and kinematic ages of hot planet hosts.}
\textbf{a-c}, The Toomre diagrams for the USP planet hosts (red, \textbf{c}), hot Jupiter hosts (blue, \textbf{a}), and hot small planet hosts (green, \textbf{b}). 
The gray dashed lines show constant values of the total Galactic velocity $V_{\rm tot}$ in steps of 50 $\rm{km} \, \rm{s}^{-1}$.
We plot the normalized number density contour lines, representing values of 0.25, 0.5, and 0.75.
In each panel, we print the kinematic ages of the corresponding host stars.
USP planet hosts exhibit a more diffuse distribution on the Toomre diagrams. 
\textbf{d}, Comparison of total velocity dispersions ($\sigma_{\rm{tot}}$) and the kinematic ages among USP planet hosts, hot Jupiter hosts, and hot small planet hosts.
The colored histograms represent the distributions of velocity dispersions obtained from 10,000 resamplings.
The colored vertical dashed lines indicate the medians of the velocity dispersion distributions.
We derive the kinematic ages using the Age-Velocity dispersion Relation (AVR, Equation \ref{Eq_S6} in Methods) represented by the black solid line.
The data points with error bars show the median values and $\pm1\sigma$ ranges of $\sigma_{\rm{tot}}$ and the kinematic ages.
USP planet hosts have larger $\sigma_{\rm{tot}}$ and are dynamically older.
}}
\end{figure*}

We first compare the kinematic properties of the host stars of USP planets, HJs, and HS planets.
\hyperlink{figure_1}{Fig. 1} depicts the cumulative distributions of the total velocity ($V_{\rm tot}$) and the relative probabilities between thick disk to thin disk ($TD/D$) for host stars of the three planetary populations. 
As evident from the data, USP planet hosts exhibit higher values of $V_{\rm tot}$ and $TD/D$ compared to HJ and HS planet hosts. 
Approximately $9.4^{+5.6}_{-3.7}\%$ $(6/64)$ of USP planet hosts are likely to be Galactic thick disk stars, assuming a threshold of $TD/D>2$.
However, this fraction diminishes to $3.8^{+1.2}_{-1.0}\%$ $(15/400)$ for HS planet hosts and further decreases to $1.4^{+3.2}_{-1.1}\%$ $(1/72)$ for HJ hosts.
We conduct a two-sample K-S test to ascertain the statistical significance of these observations.
For the $V_{\rm tot}$ distribution, the K-S test yields $p-$values of 0.0025 and 0.0359 for the comparisons between USP planet hosts and HJ hosts and between USP planet hosts and HS planet hosts, respectively. 
Similarly, for the $TD/D$ distribution, the test produces $p-$values of 0.0014 and 0.0379 for the USP-HJ and USP-HS comparisons, respectively.

The differences in the kinematic properties among these three categories of planet host stars can also be seen from the Toomre diagram, as shown in \hyperlink{figure_2}{Fig. 2a}.
As can be seen, USP planet hosts are more spread out in the velocity space, with a velocity dispersion ($\sigma_{\rm{tot}}$) of $55.27^{+0.17}_{-0.17} \, \rm{km} \, \rm{s}^{-1}$, significantly larger than those of HJ ($39.35^{+0.15}_{-0.15} \, \rm{km} \, \rm{s}^{-1}$) and HS ($46.83^{+0.06}_{-0.06} \, \rm{km} \, \rm{s}^{-1}$) planet hosts.

We then calculate the average kinematic ages of these three categories of planet host stars from $\sigma_{\rm{tot}}$ by using the Age-Velocity dispersion Relation (AVR) (see \hyperlink{4.3}{Methods \S4.3} and \hyperlink{table_s2}{Supplementary Table 2}). 
The derived ages are $5.71^{+0.72}_{-0.58}$ Gyr for USP planet hosts, $3.77^{+0.41}_{-0.34}$ Gyr for HS planet hosts, and $2.44^{+0.23}_{-0.19}$ Gyr for HJ hosts, as depicted in \hyperlink{figure_2}{Fig. 2}. 
The error bars associated with the kinematic ages are determined through the Monte Carlo method, which involves resampling the AVR coefficients and the velocity dispersion ($\sigma_{\rm{tot}}$) based on their respective uncertainties.
Out of 10,000 Monte Carlo resampling cases, the calculated kinematic ages of USP planet host stars exceed those of HJ hosts in all 10,000 cases, and they are also older than the ages of HS planet hosts in 9,978 cases. This translates to a confidence level of over 99.99\% for the statement that USP planet hosts are older than HJ hosts and 99.78\% for the statement that USP planet hosts are older than HS planet hosts. 
 During the revision process of this paper, we note that another paper was recently published by Schmidt et al. (2024) \citep{2024AJ....168..109S}, which also used the AVR method and found that USP planets are an old population with an average age of 4.7-5.8 Gyr, consistent with our derived age of $5.71^{+0.72}_{-0.58}$ Gyr.
However, as will be shown below, our research delves further into the relationship between the occurrence of USP planets and their age (\hyperlink{figure_3}{Fig. 3}), and uncovers how their architectural features vary with age (\hyperlink{figure_4}{Fig. 4}), thus providing deeper insights into their formation and evolution.

\hypertarget{R2}{\subsection*{USP planet frequency increases with time.}}
\phantomsection\label{sec:R2}

\begin{figure*}[!t]
\centering
\includegraphics[width=\textwidth]{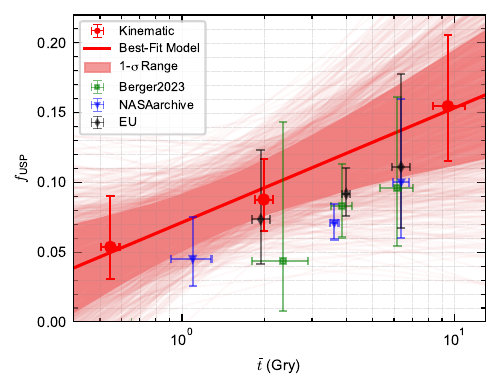}
\hypertarget{figure_3}{\caption{\textbf{The frequencies of USP planets ($f_{\rm{USP}}$) as functions of average age from different indicators.}
The data points with error bars show the median values and $\pm1\sigma$ ranges of kinematic ages on the horizontal axis, and the frequencies with Poisson counting errors on the vertical axis.
The sources for the ages are as follows:
(1) Kinematic (red dots): The kinematic ages obtained from the Age-Velocity dispersion Relation (AVR) as in previous studies of the PAST project \citep{2021ApJ...909..115C}. 
(2) Berger2023 (green squares): Isochrone ages from Berger et al. (2023) \citep{2023arXiv230111338B}. 
(3) NASAarchive (blue inverted triangles): Ages retrieved from the NASA exoplanet archive \citep[https://exoplanetarchive.ipac.caltech.edu;][]{2013PASP..125..989A}. 
(4) EU (black diamonds): Ages obtained from the Extrasolar Planets Encyclopaedia (https://exoplanet.eu). 
The red solid line and shaded region denote the best-fit linear model and 1 - $\sigma$ interval for the kinematic sample.
The best-fit linear model follows $f_{\rm USP} = 0.0821^{+0.0446}_{-0.0438} \times \log_{10}(t/\rm Gyr) + 0.0713^{+0.0197}_{-0.0202}$ (Equation \ref{Eq_1}).
The thin line of light red represents the results of 10,000 fits from resampling the kinematic sample.
}}
\end{figure*}

The comparable metallicity distributions observed in USP planet hosts and HS planet hosts hint at a potential common origin, whereas the disparity in ages between these two types of planet hosts indicates that the occurrence rate of USP planets among hot small mass planets undergoes an evolutionary process with age.

To demonstrate this, we divide our sample equally into three subsamples according to the order of $TD/D$. 
Since $TD/D$ is a proxy for relative age \citep{2021ApJ...909..115C}, these three subsamples represent younger, intermediate, and older hot small mass planet systems, whose kinematic ages are $0.55_{-0.04}^{+0.04}$ Gyr, $2.00_{-0.14}^{+0.17}$ Gyr, and $9.46_{-1.12}^{+1.46}$ Gyr, respectively, according to AVR. 
In order to isolate the effect of age, we have controlled other stellar parameters, including $T_{\rm eff}$ and $\rm [Fe/H]$, to ensure comparable distributions across the younger, intermediate, and older subsamples (see \hyperlink{Frequency}{Methods \S5} and \hyperlink{figure_s12}{Supplementary Fig. 12}).
We then calculate the relative frequency of USP planets among all hot small mass planets, i.e., $f_{\rm USP} = N_{\rm USP}/(N_{\rm USP}+N_{\rm HS})$, where $N_{\rm USP}$ and $N_{\rm HS}$ are the number of USP and HS planets respectively. 
We find that the frequency of USP planets ($f_{\rm USP}$) increases with age, which are $0.0538^{+0.0364}_{-0.0232}$ $(5/93)$, $0.0877^{+0.0290}_{-0.0224}$ $(15/171)$, and $0.1546^{+0.0511}_{-0.0395}$ $(15/97)$, for the younger, intermediate, and older subsamples, respectively.

We fit the age-$f_{\rm USP}$ relation with a simple linear function (with age in logarithm, as shown in \hyperlink{figure_3}{Fig. 3}), and the best fit is: 
\begin{equation}
f_{\rm USP} = 0.0821^{+0.0446}_{-0.0438} \times \log_{10}(t/\rm Gyr) + 0.0713^{+0.0197}_{-0.0202},
\label{Eq_1}
\end{equation}
Given the data uncertainties, the error bars of the fitting parameters are obtained by resampling and refitting the data points 10,000 times.
Among such 10,000 resampling and refitting processes, $f_{\rm USP}$ is positively correlated with age in 9,690 cases, corresponding to a confidence level of 96.90\%.

As a consistency check, we adopt individual ages measured from the literature instead of using the kinematic age above.
Specifically, we consider three ages sources: the catalog of Berger et al. (2023) \citep{2023arXiv230111338B}, NASA Exoplanet Archive \citep[\url{https://exoplanetarchive.ipac.caltech.edu};][]{2013PASP..125..989A} and the Extrasolar Planets Encyclopaedia (\url{https://exoplanet.eu}).
With the same procedure, we divide the sample into three subsamples according to age and control $T_{\rm eff}$ and $\rm [Fe/H]$ to isolate the age effect.
The derived $f_{\rm USP}$ using different age sources are presented in \hyperlink{figure_3}{Fig. 3}. 
As can be seen, a positive correlation between age and $f_{\rm USP}$, although weak, can still be observed when using other age sources, and their results align well with the one that used the kinematic age.

\hypertarget{R3}{\subsection*{USP planetary system architecture changes with time.}}
\phantomsection\label{sec:R3}

\begin{figure*}[!t]
\centering
\includegraphics[width=\textwidth]{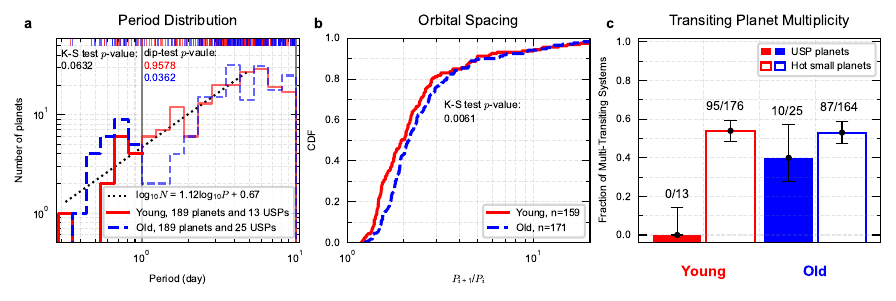}
\hypertarget{figure_4}{\caption{\textbf{Comparison of the architecture between the younger and older USP planet systems and hot small planet systems.}
\textbf{a}, Comparison of the orbital period distributions of the innermost planet between younger (red solid line) and older (blue dashed line) planetary systems. 
The two distributions differ from each other, with a K-S test $p-$value of 0.0632.
The younger distribution fits well with a single power law distribution (black dotted line), while the older one exhibits a dip-pileup feature. 
The dip test $p-$value accesses the probability that the distribution in $\log P$ for $P$ in the range 0 - 2.5 days is drawn from a unimodal distribution. 
\textbf{b}, Comparison of the neighboring period ratio ($P_{i+1}/P_{i}$) distributions in multiple systems between younger (red solid line) and older (blue dashed line) planetary systems. 
The two distributions differ significantly, with the older one shifting towards a larger period ratio. A K-S test confirms this difference, yielding a $p-$value of 0.0061.
\textbf{c}, Comparison of the fractions of multi-transiting planetary systems between younger (red histograms) and older (blue histograms) populations, and between USP planet systems (filled histograms) and hot small planet systems (open histograms).
The fractions and their Poisson counting errors for the respective bins are shown as data points with error bars.
The specific numbers used to calculate each fraction are also indicated above the corresponding histograms.
}}
\end{figure*}

We further investigate the age-dependent variations in the architecture of USP planetary systems. 
Similar to the aforementioned occurrence analysis, we divide the whole sample into subsamples and control stellar parameters to isolate the age effect (see \hyperlink{architecture}{Methods \S6} and \hyperlink{figure_s13}{Supplementary Fig. 13}).
Due to the requirement of a larger sample size for architectural analysis, our study is confined to examining only two subsamples: a younger one and an older one, distinguished by their kinematic ages of $0.69_{-0.05}^{+0.05}$ Gyr and $5.53_{-0.56}^{+0.70}$ Gyr respectively.
Our findings indicate that the younger and older subsamples exhibit distinct planetary architecture characteristics (see \hyperlink{figure_s16}{Supplementary Fig. 16} for an overview of planetary architecture in the younger and older subsamples).
Specifically, in the following, we highlight the discrepancies in three aspects: orbital period, orbital spacing, and transiting planet multiplicity.

\emph{Period distribution.} 
\hyperlink{figure_4}{Fig. 4a} displays the period distribution of the innermost planets in hot planet systems.
For the younger subsample, the period distribution aligns well with a power-law distribution, with a K-S test $p-$value of 0.9792. 
In contrast, the period distribution of the older subsample deviates from that of the younger subsample, with a K-S test $p-$value of 0.0632, and also from a power-law distribution, with a K-S test $p-$value of 0.1410.
Furthermore, a notable dip/pileup feature emerges just beyond/within $period=1$ day, particularly for the older subsample.
The Hartigan's dip tests return a $p-$value of 0.9578 (i.e., a confidence level of 95.78\% supporting the absence of a dip feature) for the younger subsample and a $p-$value of 0.0362 (i.e., a confidence level of $1-0.0362=96.38\%$ supporting the presence of a dip feature) for the older subsample, respectively.
Such an age-dependent period distribution not only confirms the aforementioned finding that the occurrence of USP planets increases with age, but also illustrates how USP planets emerge over time.

\emph{Orbital spacing.} 
\hyperlink{figure_4}{Fig. 4b} presents the orbital spacing (in terms of the orbital period ratio of adjacent planets) distribution of hot multiple planet systems in our sample. 
Notably, the older subsample exhibits a shift towards a larger period ratio (with a median of 2.13) compared to the younger subsample (with a median of 1.94). 
The KS test of these two period ratio distributions yields a $p-$value of 0.0061, indicating that the two distributions are distinct at a confidence level of $1-0.0061=99.39\%$. 
It is noteworthy that the difference in period ratio is observed for all adjacent pairs, not just the innermost ones (see \hyperlink{figure_s17}{Supplementary Fig. 17}-\hyperlink{figure_s18}{18}). 

\emph{Transiting planet multiplicity.} 
\hyperlink{figure_4}{Fig. 4c} compares the transiting planet multiplicities of the younger and older subsamples of USP and hot small planet systems.
As can be seen, USP planet systems generally have a lower fraction of multiple transiting systems than hot small planet systems \citep{2013ApJ...774L..12S,2016PNAS..11312023S}.
For hot small planet systems, 95 out of 176 ($\sim53.98^{+5.54}_{-5.54}\%$) systems in the younger subsample exhibit multiple transiting planets, which is comparable to that of the older subsample ($87/164, \sim53.05^{+5.69}_{-5.69}\%$).
In contrast, for the USP planet systems, the fraction of multiple transiting systems differs significantly between the younger ($0/13, \sim0.00^{+14.16}\%$) and older ($10/25, \sim40.00^{+17.06}_{-12.43}\%$) subsamples. 
This implies that younger USP planets may have fewer nearby companions and/or higher mutual inclinations.
This result is further supported by additional complementary evidence obtained from examining the fraction of non-transiting planets.
We find that 2 out of the 13 ($\sim15.38^{+20.29}_{-9.94}\%$) USP planet systems in the younger subsample have non-transiting planetary companions, while none are found among the USP planets systems ($0/25, \sim0.00^{+7.36}\%$) in the older subsample (see \hyperlink{figure_s16}{Supplementary Fig. 16}). 
In other words, if focusing solely on multiple planet systems, in the younger USP planet systems, all (2 out of 2) multiple planet systems are systems with non-transiting planets.
Conversely, in the older USP planet systems, all (10 out of 10) multiple planet systems are systems without non-transiting planets.
However, we caution that the current sample of non-transiting companions in USP planet systems is small, with only two observations coming from different surveys, namely CoRoT and K2, and thus may be subject to small number statistics and potential observational biases.

The above results are based on a heterogeneous sample, in which USP planets were discovered by {\it Kepler}, K2, TESS, and CoRoT.
To account for the effect of sample selection and survey completeness, we repeat the above analyses focusing on the {\it Kepler} sample (see \hyperlink{Supplementary_discussion}{Methods \S7}, supplementary discussion), which comprises approximately two-thirds (43/64) of the USP planets in the total sample.
As shown in \hyperlink{figure_s19}{Supplementary Fig. 19a} and \hyperlink{figure_s20}{Supplementary Fig. 20}, the planet detection efficiencies around {\it Kepler} stars of different ages are comparable.
Moreover, we still find that in the {\it Kepler} sample, the frequency of USP planets increases with age (\hyperlink{figure_s19}{Supplementary Fig. 19}), and the architecture of USP planetary systems varies with age (\hyperlink{figure_s21}{Supplementary Fig. 21}). 
The results from the {\it Kepler} sample are generally consistent with the aforementioned findings but have slightly lower confidence levels and larger error bars due to the reduced sample size.

\hypertarget{Discussions_and_conclusions}{\section*{Discussions and conclusions}}
\phantomsection\label{sec:Discussions_and_conclusions}

Our results have substantial implications for understanding the formation and evolution of USP planet systems. 
The observed trend, wherein USP planet frequency increases with time (\hyperlink{figure_3}{Fig. 3}), suggests that the majority of USP planets likely form after $\sim$1 Gyr. 
An earlier formation would result in a flat or even declining USP frequency over time, due to orbital decay caused by magnetic effects at early stages, as proposed by Strugarek et al. (2017) \citep{2017ApJ...847L..16S}, which contradicts our observations.
This argues against very early formation models, such as inward migration in protoplanetary disks \citep{2021ApJ...919...76B}, as the primary channel for USP planet formation.

Instead, given the slow nature of tidal effects, formation models involving inward migration driven by tidal dissipation are more plausible.
Considering various sources of tidal dissipation, several tidal mechanisms have been proposed, including stellar tides \citep{2010ApJ...724L..53S,2017ApJ...842...40L}, planetary eccentricity tides (which can be divided into a high-eccentricity model \citep{2019AJ....157..180P} and a low-eccentricity model \citep{2019MNRAS.488.3568P}), and planetary obliquity tides \citep{2020ApJ...905...71M}.
Our findings regarding the age dependence of USP planetary architecture (\hyperlink{figure_4}{Fig. 4}) provide observational evidence for certain predictions made by tidal models, offering new insights into the formation and evolution of USP planets.

Firstly, the observed age-dependent period distribution, particularly the dip-pileup feature (\hyperlink{figure_4}{Fig. 4a}), may indicate that the migration processes leading to the formation of USP planets are not smooth.
This result is not unexpected in light of the aforementioned tidal models. 
On one hand, the low-eccentricity migration model has the potential to form a dip feature in the period distribution \citep[e.g., Figure 16 of][]{2019MNRAS.488.3568P}, given certain suitable initial orbital periods and eccentricities.
On the other hand, the high-eccentricity migration model can cause planets to undergo a runaway inward migration, typically stalling at a period of $\sim 1$ days \citep[i.e., tidal capture locations, e.g., Equation 20 of][]{2019AJ....157..180P},
which resembles the observed pileup.
Besides planetary tides, another potential explanation for the dip-pileup in the period distribution could be tidal migrations driven by stellar tides \cite{2017ApJ...842...40L}.
However, this hypothesis requires that the tidal dissipation (i.e., the tidal $Q$ factor) within the star varies with the orbital period. 
Although there is some evidence suggesting period-dependent tidal dissipation in hot Jupiter systems \citep{2018AJ....155..165P}, it remains uncertain whether this also applies to USP planet systems. 
Furthermore, even if such period-dependence exists in USP planet systems, the specific form of the tidal $Q$ factor's dependence required to reproduce the observed dip-pileup in the period distribution appears to be ad-hoc and not yet well-understood \citep{2007ApJ...661.1180O,2019MNRAS.484.3017W,2024ApJ...963...34W}. 
Additionally, the position of the dip-pileup may reinforce the notion that a period of 1-2 days serves as the boundary between USP planets and other short-period planets \citep{2025AJ....169..191G}.

Secondly, as tidal orbital decay is generally more pronounced for closer-in planets, it is reasonable to expect that the orbital periods of the inner planets would drift faster towards the star, and consequently, the orbital spacings between adjacent planets would increase over time. 
Indeed, such effects are evident in our results, as depicted in \hyperlink{figure_4}{Fig. 4a,b}.
However, it appears unlikely that tidal effects alone can fully account for the wider orbital spacing observed in older systems, as this feature is not only present in the innermost planet pairs but also in the outer pairs (see \hyperlink{figure_s17}{Supplementary Fig. 17}-\hyperlink{figure_s18}{18}), where tidal effects are weak.
Therefore, apart from tides, additional mechanisms must be invoked to explain the widening of planetary orbital spacing over timescales of billions of years. 

Thirdly, USP planet formation often involves exciting orbital inclinations \citep{2019AJ....157..180P,2019MNRAS.488.3568P,2020ApJ...905...71M}.
In fact, the larger mutual inclinations of USP planets as compared to other hot small planets with longer periods were revealed from observations \citep{2018ApJ...864L..38D} before most of the theories proposed.
Our result (\hyperlink{figure_4}{Fig. 4c}) shows that younger USP planet systems have significantly fewer multiple transiting planets, indicating fewer nearby companions and/or larger mutual orbital inclinations compared to older systems.
This is expected in the high-eccentricity model \citep{2019AJ....157..180P}, which typically generates USP planets with larger mutual inclinations and fewer nearby companions. 
The process is enhanced and accelerated by larger initial mutual inclinations (thus larger initial Angular Momentum Deficit, AMD), thereby explaining why these features are more prevalent in younger USP planet systems.
Alternatively, inclinations may be excited post-migration through secular perturbations from stellar oblateness, which decreases with stellar spin-down over age, making this mechanism efficient primarily for younger systems of age within 1 Gyr \citep{2020ApJ...890L..31L,2020AJ....160..254B,2021AJ....162..242B}, also consistent with our findings.

In conclusion, our work uncovers the age dependence of the occurrence and architecture of USP planetary systems, providing constraints on their formation and evolution.
The age dependence of their occurrence indicates that the formation of USP planets is a continuous process extending over billions of years, with the formation rate consistently increasing over time.
The age dependence observed in their architectural features suggests that USP planets originated through inward migration driven by tidal dissipation.
Concerning the specific tidal migration mechanisms, a combination of both the high-eccentricity migration model \citep{2019AJ....157..180P} and the low-eccentricity migration model \citep{2019MNRAS.488.3568P} seems to best fit the observed architectural characteristics.
In particular, the dip-pileup at around 1 day and the higher fraction of multi-transiting systems in the older population strongly suggest that these planets have migrated inward from nearby orbits (i.e., 1-2 day orbits) via low-eccentricity migration \citep{2019MNRAS.488.3568P}. 
Conversely, the lower fraction of multi-transiting systems in the younger population aligns with the notion that they have migrated from more distant orbits via high-eccentricity migration \citep{2019AJ....157..180P}.

Future synthetic studies, incorporating detailed simulations and comparisons with observational data, will further quantify the relative contributions of these two USP formation models.
Ongoing and future observations by missions such as TESS \citep{2015JATIS...1a4003R} and PLATO \citep{2014ExA....38..249R} will discover more USP planets across a wider age range, enabling the validation of our current findings and promoting a more profound understanding of USP planet formation and evolution.

\section*{Data availability}

\small The datasets generated and analyzed in this study are available via Zenodo at \url{https://zenodo.org/records/14221205} \citep{tu_2025_14221205}.
Other data used in this study are accessible from the public archives of the observatories or survey websites.

\section*{Code availability}

\small The KeplerPORTs software package\citep{2015ApJ...809....8B,2017ksci.rept...19B} is publicly available at \url{https://github.com/nasa/KeplerPORTs.}

\section*{Acknowledgements}

\small This work is supported by the National Key R\&D Program of China (No. 2024YFA1611803) and the National Natural Science Foundation of China (NSFC; Grant Nos. 12273011, 12150009, 12403071). We also acknowledge the science research grants from the China Manned Space Project with No. CMS-CSST-2021-B12. J.-W.X. also acknowledges the support from the National Youth Talent Support Program. D.-C.C. also acknowledges the Cultivation project for LAMOST Scientific Payoff, Research Achievement of CAMS-CAS, and the fellowship of Chinese Postdoctoral Science Foundation (2022M711566). Funding for LAMOST (www.lamost.org) has been provided by the Chinese NDRC. LAMOST is operated and managed by the National Astronomical Observatories, CAS.

\section*{Author contributions statement}

\small J.-W.X. conceived the project and designed the research; P.-W.T. led the data analyses; P.-W.T., J.-W.X., and D.-C.C. analyzed the results and drafted the manuscript; and all authors contributed to discussing the results, editing, and revising the manuscript.

\section*{Competing interests statement}

\small The authors declare no competing interests.

\clearpage

\section*{Methods \& Supplementary Material} 

\hypertarget{Data_sample}{\subsection*{1. Data sample}}
\phantomsection\label{sec:Data_sample}

\hypertarget{1.1}{\subsubsection*{1.1 Sample selection}}
\phantomsection\label{sec:1.1}

\hypertarget{1.1.1}{\subsubsection*{1.1.1 Obtaining the hot planet host sample}}
\phantomsection\label{sec:1.1.1}

We initialize our sample of hot planets ($P < 10$ days) discovered by transit method around Sun-like stars with effective temperature ($T_{\rm{eff}}$) in the range of $4700-6500$ ${\rm K}$ and surface gravity ($\log g$) $>4.0$ from two sources:

\begin{enumerate}

\item {\it Kepler} DR25: 
{\it Kepler} Data Release 25 \citep{2018ApJS..235...38T} has identified 6923 host stars and 8054 {\it Kepler} Objects of Interest (KOIs). 
After excluding false positives identified in {\it Kepler} DR25, 4,034 planet candidates around 3,069 stars remain.
It is worth noting that additional motion caused by binary orbits may affect the results of subsequent kinematic characterization \citep{1998MNRAS.298..387D,2021ApJ...909..115C}.
Therefore, we exclude stars with “$cb\_flag$” $\neq 0$ and “$sy\_snum$” $> 1$ in the stellar host catalog of the NASA Exoplanet Archive \citep[\url{https://exoplanetarchive.ipac.caltech.edu},][EA hereafter]{2013PASP..125..989A} as well as in the binary catalog of the Extrasolar Planets Encyclopaedia (\url{https://exoplanet.eu}, EU hereafter) to remove potential binaries, leaving 3,943 confirmed planets around 3,019 stars.
We further refine this sample by selecting only those with transit signal-to-noise ratios (S/N) greater than 7.1 \citep{2011ApJ...736...19B} and disposition scores above 0.9 \citep{2018ApJS..235...38T}, which results in 3,345 planet candidates around 2,582 stars.
We crossmatch these host stars with Berger et al. (2020) \citep{2020AJ....159..280B} to obtain homogeneous stellar properties such as $T_{\rm{eff}}$, $\log g$, and stellar radius ($R_{\rm{star}}$). 
Further, we have 3,186 planet candidates around 2,450 stars. 
We calculate the radius and its uncertainty for each planet based on the planet-to-star radius ratios (roR) provided by {\it Kepler} DR25 and the stellar radii from Berger et al. (2020). 
We further narrow down our selection to planets with a radius less than 20 Earth radii and a radius uncertainty-to-radius ratio smaller than 0.5.
This selection results in 3,092 planet candidates around 2,366 stars. 
Finally, we focus on hot planets around Sun-like stars, leaving 1,015 {\it Kepler} stars hosting 1,147 planets. 

\item Confirmed Planets: EA has provided 4,236 host stars with 5,678 confirmed planets (by 6/25/2024). 
We also remove potential binaries listed in the stellar host catalog of EA and the binary catalog of EU to minimize the effect of binary orbits on kinematic characterization, leaving 5,138 confirmed planets around 3,839 stars.
For these stars, we adopt stellar properties ($T_{\rm{eff}}$, $\log g$) from literature in the EA. It is worth noting that these stellar properties are inhomogeneous. 
We also apply a cut on the relative error of $R_{\rm{planet}}$ and exclude planets with a radius larger than 20 Earth radii, similar to the process used in the {\it Kepler} sample.
Consequently, we have 3,463 confirmed planets around 2,630 stars.
We remove the planets already included in the {\it Kepler} sample. 
In this section, we obtain 574 Sun-like stars hosting 627 transiting hot planets. 

\end{enumerate}

We combine the samples from the two aforementioned sources. 
This combined planet host sample contains 1,801 transiting hot planets around 1,589 stars. 
The sample selection process is detailed in \hyperlink{table_s1}{Supplementary Table 1} for reading convenience.

\begin{table}[!t]
\renewcommand\arraystretch{1.5}
\centering
\hypertarget{table_s1}{{\bfseries Supplementary Table 1: \\Rundown table of sample selection: the transiting hot planets around Sun-like stars.}}
{\footnotesize
\begin{tabular}{l|cc|cc|cc} \hline
\hline
 & \multicolumn{2}{c}{Kepler DR25}&\multicolumn{2}{c}{Other Confirmed}&\multicolumn{2}{c}{Sum}\\ \hline
&$N_{\rm s}$&$N_{\rm p}$&$N_{\rm s}$&$N_{\rm p}$&$N_{\rm s}$&$N_{\rm p}$\\ \hline
\multicolumn{7}{c}{Section 1.1.1 Obtaining the Planet Host Sample}\\ \hline
Initial&6923&8054&4236&5678&-&-\\
Not false positive and binary&3019&3943&3839&5138&-&-\\
S/N $>$ 7.1 and disposition score $>$ 0.9&2582&3345&3839&5138&-&-\\
GOF $>$ 0.99 \citep{2020AJ....159..280B}&2450&3186&3839&5138&-&-\\
$R_{\rm p}$ $<$ 20 $R_{\rm earth}$ and relative error of $R_{\rm p}$ $<$ 0.5&2366&3092&2630&3463&-&-\\
Removal of duplication&2366&3092&1288&1561&3654&4653\\
Main sequence $\log g > 4.0$ dex&2181&2884&1142&1389&3323&4273\\
$4700 {\rm K} < T_{\rm eff} < 6500{\rm K}$&1861&2449&853&1023&2704&3472\\
$P < 10 $ days&1015&1174&574&627&1589&1801\\
\hline
\multicolumn{7}{c}{Section 1.1.2 Obtaining Astrometric Parameters}\\
\hline
Relative error of parallax $< 0.1$ and $\rm{RUWE} < 1.2$&923&1067&509&551&1432&1618\\
\hline
\multicolumn{7}{c}{Section 1.1.3 Obtaining Spectroscopic Parameters}\\
\hline
With LAMOST&251&298&62&73&313&371\\
With CKS&411&524&16&23&427&547\\
With Gaia&289&348&399&435&688&783\\
\hline
\multicolumn{7}{c}{Section 1.2 Calibrating Parameters from Various Sources}\\
\hline
Combined LAMOST, CKS, and Gaia&501&624&408&447&909&1071\\
\hline
\multicolumn{7}{c}{Section 1.3 Classifying Galactic Composition}\\
\hline
Stars in Galactic disk&458&572&368&403&826&975\\
\hline
\end{tabular}}
\flushleft
{\centering $N_{\rm s}$ and $N_{\rm p}$ are the numbers of host stars and planets during the process of sample selection in \hyperlink{Data_sample}{Methods \S1}.}
\end{table}

\hypertarget{1.1.2}{\subsubsection*{1.1.2 Obtaining astrometric parameters}}
\phantomsection\label{sec:1.1.2}

We then crossmatch our transiting hot planet hosts sample with the third Gaia data release \citep[Gaia DR3,][]{2023A&A...674A...1G} via the Gaia Archive (\url{https://gea.esac.esa.int/archive/}) to obtain five astrometric parameters for each star: positions ($\alpha$ and $\delta$), parallaxes, proper motions ($\mu_{\alpha}$ and $\mu_{\delta}$).
The crossmatching is limited to an angular separation of less than 1.5 arcseconds and a G-band magnitude difference of less than 2, to ensure similar position and brightness. 
If more than one source meets the matching criteria, we retain the one with the smallest angular separation.
We constrain the relative error of parallax to be below 0.1 to ensure the accuracy of the astrometric parameters \citep{1998MNRAS.298..387D}.
Additionally, we further remove stars with Gaia DR3 $\rm{RUWE}$ (re-normalized unit-weight error) greater than 1.2, as they are likely to be potential binaries \citep{2018AJ....156..195R,2020AJ....159..280B}. 
Finally, we obtain 1,618 transiting hot planets around 1,432 stars with Gaia astrometric parameters.

\hypertarget{1.1.3}{\subsubsection*{1.1.3 Obtaining spectroscopic parameters}}
\phantomsection\label{sec:1.1.3}

We also obtain additional spectral parameters for our transiting hot planet hosts sample using three sources: LAMOST DR9, CKS, and Gaia DR3.

\begin{enumerate}

\item LAMOST: The Large Sky Area Multi-Object Fiber Spectroscopic Telescope (LAMOST) contains tens of millions of stellar spectra \citep{2012RAA....12.1197C}.
The LAMOST DR9 v1.0 low-resolution catalog (\url{https://www.lamost.org/dr9/v1.0/catalogue}) provides 7.06 million stellar spectra for A, F, G, and K stars. 
The typical uncertainties for RVs and $\rm [Fe/H]$ are 5.0 $\rm{km} \, \rm{s}^{-1}$ and 0.04 dex, respectively.
We crossmatch the LAMOST data with our transiting hot planet host sample to obtain RV and $\rm [Fe/H]$.
We apply a quality cut of S/N $>$ 15 and ensure that the ratio of RV to its uncertainty is greater than 3. 
As a result, we obtain 371 hot planets around 313 stars. 

\item CKS: The California-Kepler Survey (CKS) project measures the precise properties of planets and their host stars discovered by NASA's {\it Kepler} Mission. 
It provides high-resolution spectra for 1,305 stars hosting {\it Kepler} transiting planets, including radial velocity (RV) and chemical abundance ($\rm [Fe/H]$) \citep{2017AJ....154..107P}. 
The RVs and $\rm [Fe/H]$ have precisions of 0.1 $\rm{km} \, \rm{s}^{-1}$ and 0.04 dex, respectively. 
The quality control cut requires the ratio of RV to its uncertainty to be greater than 3. 
We crossmatch our transiting hot planet host sample with the CKS data, resulting in 547 planets around 427 {\it Kepler} stars. 

\item Gaia: Gaia DR3 contains the radial velocities of more than 33 million stars \citep{2023A&A...674A...5K}.
The typical uncertainties of RVs are 1.3 $\rm{km} \, \rm{s}^{-1}$ at $G_{\rm RVS} = 12$, and 6.4 $\rm{km} \, \rm{s}^{-1}$ at $G_{\rm RVS} = 14$.
Applying a quality cut of S/N ({\it rv\_expected\_sig\_to\_noise}) greater than 5 and a ratio of RV to its uncertainty exceeding 3, we crossmatch the transiting hot planet host sample to obtain RV. 
This process identifies 783 hot planets around 688 stars.

\end{enumerate}

Finally, we obtain 1,071 hot planets around 909 stars with spectroscopic parameters from at least one of these sources.

\hypertarget{1.2}{\subsubsection*{1.2 Calibrating parameters from various sources}}
\phantomsection\label{sec:1.2}

As shown in the aforementioned section, we obtain stellar spectral parameters from LAMOST DR9, CKS, and Gaia DR3. 
Systematic parameter differences from different sources can lead to biases in our statistical analysis. 
Therefore, it is necessary to calibrate these systematic differences across various sources. 

We crossmatch the CKS high-resolution spectra \citep{2017AJ....154..107P} used in the aforementioned section with both the LAMOST DR9 v1.0 low-resolution catalog and Gaia DR3. 
We apply the same quality cuts to the spectral data from each source as those used in the aforementioned section. 
Additionally, we retain Sun-like stars based on $\log g$ and $T_{\rm{eff}}$ from CKS.
For calibration purposes, we use the RVs from CKS and $\rm [Fe/H]$ from LAMOST as standards for calibration.
To eliminate measurements with large offsets, we apply the median absolute deviation (MAD) as a criterion to remove measurements where the differences in RV or $\rm [Fe/H]$ are larger than 5 MAD. 

Based on the standards of each parameter, we use the best linear fit function between the parameter to be calibrated and the standard parameter as the calibration function. 
The function is as follows:

\begin{equation}
\begin{split}
{\rm [Fe/H]}_{\rm LAMOST} = ({\rm [Fe/H]}_{\rm CKS} - 0.021) \div 0.881 \\
{\rm RV}_{\rm CKS} = ({\rm RV}_{\rm LAMOST} + 6.160) \div 0.981 \\
{\rm RV}_{\rm CKS} = ({\rm RV}_{\rm Gaia} - 0.172) \div 0.999 
\end{split}
\label{Eq_S1}
\end{equation}

The comparison of parameters and the fitting results are shown in \hyperlink{figure_s1}{Supplementary Fig. 1}. 
The final uncertainty of the calibrated parameters is given by $\sqrt{\sigma_{o}+\sigma_{s}}$, where $\sigma_{o}$ is the original uncertainty provided by the source of the parameter, and $\sigma_{s}$ is the standard deviation of the corresponding fitting residuals, as shown in \hyperlink{figure_s1}{Supplementary Fig. 1}. 
For ${\rm [Fe/H]}_{\rm CKS}$, $\sigma_{s}$ is 0.04 ${\rm dex}$; for ${\rm RV}_{\rm LAMOST}$, it is 3.70 $\rm{km} \, \rm{s}^{-1}$; and for ${\rm RV}_{\rm Gaia}$, it is 1.58 $\rm{km} \, \rm{s}^{-1}$.
For stars with multiple RV sources, the precedence order follows the precision order: CKS, followed by Gaia, and finally LAMOST.
For stars with multiple sources of $\rm [Fe/H]$, the order of precedence is LAMOST, followed by CKS.
Both LAMOST and CKS provide $\rm [Fe/H]$ measurements with comparable precisions.
We prefer the LAMOST-based $\rm [Fe/H]$ because it allows us to independently test the metallicity trend of USP planets, as found by Winn et al. (2017) \citep{2017AJ....154...60W}, which adopted $\rm [Fe/H]$ measurements from CKS.

\begin{figure}[!t]
\centerline{\includegraphics[width=\textwidth]{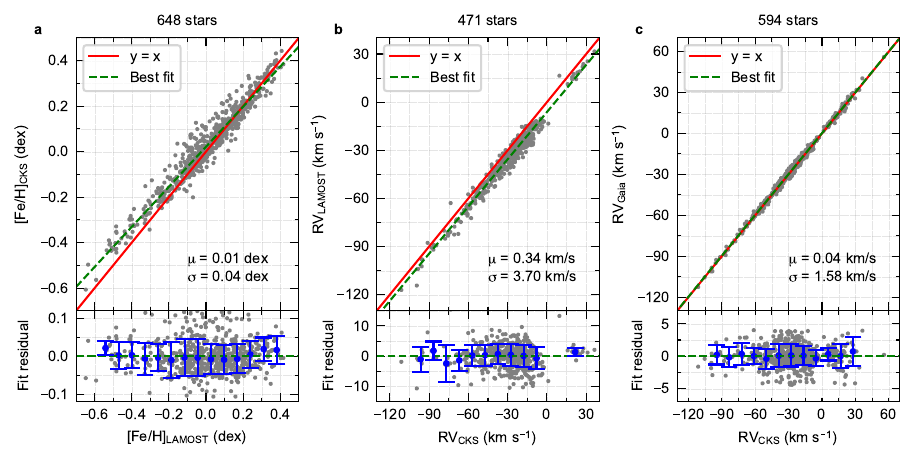}}
{\textbf{\hypertarget{figure_s1}{Supplementary Fig. 1: Calibration of parameters from various sources.}} 
Panels show the comparisons between ${\rm [Fe/H]}_{\rm LAMOST}$ and ${\rm [Fe/H]}_{\rm CKS}$ (\textbf{a}), ${\rm RV}_{\rm CKS}$ vs. ${\rm RV}_{\rm LAMOST}$ (\textbf{b}), and ${\rm RV}_{\rm CKS}$ vs. ${\rm RV}_{\rm Gaia}$ (\textbf{c}). 
The red solid lines indicate where the horizontal and vertical coordinates are equal. 
The green dashed lines represent the best linear fits of the horizontal values and vertical values, with the values of the median ($\mu$) and standard deviation ($\sigma$) of the fitting residuals marked in the bottom-right corner. 
The lower part of each panel shows the fitting residuals. 
The blue dots and error bars in each panel represent the median and standard deviation of the residuals in the individual bins.}
\end{figure}

\hypertarget{1.3}{\subsubsection*{1.3 Classifying galactic composition}}
\phantomsection\label{sec:1.3}

We briefly describe how we classify stars into different Galactic components.
We calculate the 3D Galactocentric cylindrical coordinates ($R$, $\theta$, $Z$) for the stars in our sample by adopting the Sun's location at $R_{\rm sun} = 8.34$ kpc \citep{2014ApJ...783..130R} and $Z_{\rm sun} = 27$ pc \citep{2001ApJ...553..184C}. 
To calculate the Galactic rectangular velocities relative to the local standard of rest ($U_{\rm LSR}$, $V_{\rm LSR}$, $W_{\rm LSR}$) and their errors for each star, we adopt the formulae and matrix equations from Johnson \& Soderblom (1987) \citep{1987AJ.....93..864J} and use the solar peculiar motion values [$U_{\rm sun}$, $V_{\rm sun}$, $W_{\rm sun}$] = [9.58, 10.52, 7.01] $\rm{km} \, \rm{s}^{-1}$ \citep{2015ApJ...809..145T}. 

We adopt the revised kinematic characteristics and calculate the relative probabilities between two different components for each star based on their Galactic positions and velocities \citep[more details in][]{2021ApJ...909..115C}, i.e., the thick-disk-to-thin-disk ($TD/D$), thick-disk to halo ($TD/H$), Hercules-to-thin-disk ($Herc/D$), and Hercules-to-thick-disk ($Herc/TD$). 
Then, we classify them into different Galactic components by adopting the same criteria as Bensby et al. (2014) \citep{2014A&A...562A..71B}:

(1) thin disk: $TD/D < 0.5$ \& $Herc/D < 0.5$;

(2) thick disk: $TD/D > 2$ \& $TD/H > 1$ \& $Herc/TD < 0.5$;

(3) halo: $TD/D > 2$ \& $TD/H < 1$ \& $Herc/TD < 0.5$;

(4) Hercules: $Herc/D < 1$ \& $Herc/TD > 1$.

As the age velocity relation is only applicable to stars in the Galactic disk \citep{2021ApJ...909..115C}, we remove the stars that are in the halo and Hercules stream, keeping 826 stars in the Galactic disk hosting 975 hot planets.

\hypertarget{1.4}{\subsubsection*{1.4 Final samples}}
\phantomsection\label{sec:1.4}

We then classify the planet host sample from the Galactic disk into three categories based on the orbital period and radius of the planets for subsequent analysis and discussion:

\begin{enumerate}

\item Ultra-short-period (USP) Planets: Planets with orbital periods shorter than 1 day and radii smaller than 6 Earth radii.
According to this criterion, there are 64 USP planets in the sample.
\item Hot Jupiters (HJs): Planets with orbital periods shorter than 10 days and radii greater than 6 Earth radii. There are 225 HJs in the sample. 
\item Hot Small (HS) Planets: Planets with orbital periods in the range of 1 - 10 days and radii smaller than 6 Earth radii. There are 686 HS planets in the sample. 
\end{enumerate}

\hyperlink{figure_s2}{Supplementary Fig. 2b} displays the distribution of orbital periods and radii for the three categories of planets in the final sample.
\hyperlink{figure_s3}{Supplementary Fig. 3} depicts the distribution of $T_{\rm eff}$ and $\log g$ for the three categories of planetary host stars. 

Notably, some well-known USP planet systems are not included in our sample due to our sample selection criteria:

(1) We exclude the USP planets in binary star systems, e.g., 55 Cancri e \citep{2008ApJ...675..790F,2011ApJ...737L..18W} and K2-266 b \citep{2018AJ....156..245R}.

(2) We remove TOI-500 b because its host star \citep[$4440\pm100$K,][]{2022NatAs...6..736S} is not within the range of Sun-like stars ($4700-6500$ ${\rm K}$). 

(3) Two USP planets, WASP-47 e \citep{2015ApJ...812L..18B} and EPIC-206024342 b \citep{2021PSJ.....2..152A}, are removed because their host stars have $\rm{RUWE}$ measurements greater than 1.2, indicating that they might be in potential binary systems.

\begin{figure}[!t]
\centerline{\includegraphics[width=\textwidth]{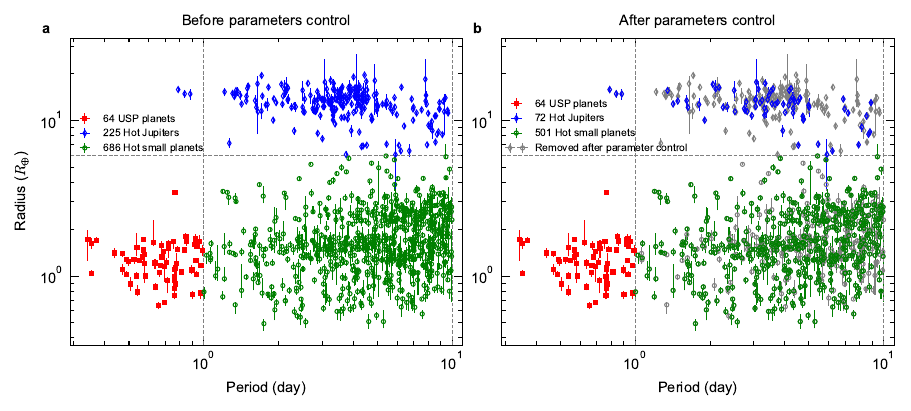}}
{\textbf{\hypertarget{figure_s2}{Supplementary Fig. 2: Orbital period and planetary radius.}}
The data points with radius uncertainties represent USP planets (red solid squares), hot Jupiters (blue open diamonds), and hot small planets (green open dots).
\textbf{a}, The samples before parameter control. 
\textbf{b}, The samples after parameter control in $T_{\rm eff}$ - $\log g$ plane. 
The gray diamonds and dots represent samples that were removed during parameter control.}
\end{figure}

\begin{figure*}[!t]
\centerline{\includegraphics[width=\textwidth]{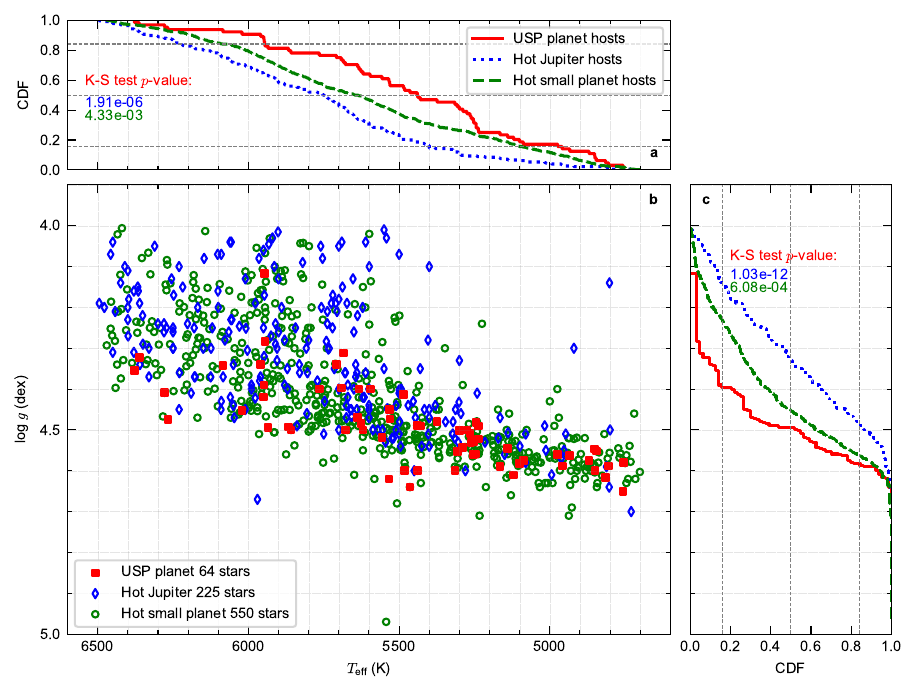}}
{\textbf{\hypertarget{figure_s3}{Supplementary Fig. 3: Comparison of $T_{\rm{eff}}$ and $\log g$ of hot planet hosts before applying the stellar parameter control.}} 
\textbf{b}, The distributions of $T_{\rm{eff}}$ and $\log g$ for the USP planet hosts (red solid squares), hot Jupiter hosts (blue open diamonds), and hot small planet hosts (green open dots). 
\textbf{a} and \textbf{c}, The cumulative distributions of $T_{\rm{eff}}$ and $\log g$, respectively. 
The horizontal gray dashed lines represent the 50$\pm$34.1 percentiles in the distribution. 
The $T_{\rm{eff}}$ and $\log g$ distributions for USP planet hosts exhibit significant differences with those of hot Jupiter and hot small planet hosts in light of typical uncertainties ($\sim96$ K for $T_{\rm{eff}}$ and $\sim0.04$ dex for $\log g$), with median values of $T_{\rm{eff}}$ at 5439 K, 5750 K, and 5642 K, and median values of $\log g$ at 4.50 dex, 4.34 dex, and 4.45 dex, respectively. 
Each panel shows K-S test $p$-values comparing hot Jupiter and hot small planet hosts to USP planet hosts.
All the K-S test $p-$values are smaller than 0.05.}
\end{figure*}

\hypertarget{Stellar_parameter_control}{\subsection*{2. Stellar parameter control}}
\phantomsection\label{sec:Stellar_parameter_control}

We need to isolate the effects of age by controlling stellar parameters (e.g., $T_{\rm eff}$, $\log g$, and $\rm [Fe/H]$). 
Our method for controlling these stellar parameters is as follows:

Step 1: In the total sample, we apply Z-score normalization separately to each parameter that needs to be controlled. 
For example, the relationship between $T_{\rm eff}$ and its Z-score normalized value ($\widehat{T_{\rm{eff}}}$) is as follows:

\begin{equation}
\widehat{T_{\rm{eff}}} = \frac{T_{\rm{eff}}-\overline{T_{\rm{eff}}}}{\sigma_{T_{\rm{eff}}}},
\label{Eq_S2}
\end{equation}
where $\overline{T_{\rm{eff}}}$ and $\sigma_{T_{\rm{eff}}}$ are the average value and standard deviation of $T_{\rm eff}$ in the total sample, respectively.

Step 2: We define the weighted Euclidean distance (d) between host stars within the parameter space after Z-score normalization as:

\begin{equation}
d = [w_{1}(\Delta \widehat{T_{\rm{eff}}})^2 + w_{2}(\Delta \widehat{\log g})^2 + w_{3}(\Delta \widehat{{\rm [Fe/H]}})^2]^{\frac{1}{2}},
\label{Eq_S3}
\end{equation}
where $w_{1}$, $w_{2}$, and $w_{3}$ are the weighting coefficients. 

Step 3: We calculate the weighted Euclidean distance ($d$) between host stars within the parameter space after Z-score normalization.

Taking a specific subsample as the standard sample, we select the nearest $n$ neighbor host stars from another sample for each host star in the standard sample.
Neighbor host stars selected multiple times will only be recorded once. 
We try different values of $w_{1}$, $w_{2}$, and $w_{3}$ (ranging from 0.0 to 1.0) and $n$ (ranging from 1 to 10) and perform a two-sample Kolmogorov-Smirnov (K-S) test between the standard sample and the selected neighbor sample for each controlled stellar parameter. 
We adopt the selected neighbor sample where the K-S test $p-$values for each controlled stellar parameter between the standard sample and the selected neighbor sample are all greater than 0.32 (a difference of less than $1 \sigma$), and where the selected neighbor sample contains the maximum number of stars.
If we cannot find suitable values for $w_{1}$, $w_{2}$, $w_{3}$, and $n$ that result in all K-S test $p-$values exceeding 0.32, we will lower the threshold to $p-$values greater than 0.05 and repeat the process. 

For example, before applying stellar parameter control, significant differences exist in the distributions of $T_{\rm{eff}}$ and $\log g$ among the host stars of the three categories of planets (see \hyperlink{figure_s3}{Supplementary Fig. 3}, all K-S test $p-$values are less than $0.01$). 
Following the steps above, we take USP planet host stars as the standard sample and apply stellar parameter control to HJ hosts and HS planet hosts in the $T_{\rm{eff}}$-$\log g$ plane ($w_{3}=0$). 
After applying stellar parameter control, we select 72 hot Jupiter hosts and 400 hot small planet hosts, whose distributions in the $T_{\rm eff}$ - $\log g$ plane are similar to those of USP planet hosts (see \hyperlink{figure_s4}{Supplementary Fig. 4}). 

\begin{figure*}[!t]
\centerline{\includegraphics[width=\textwidth]{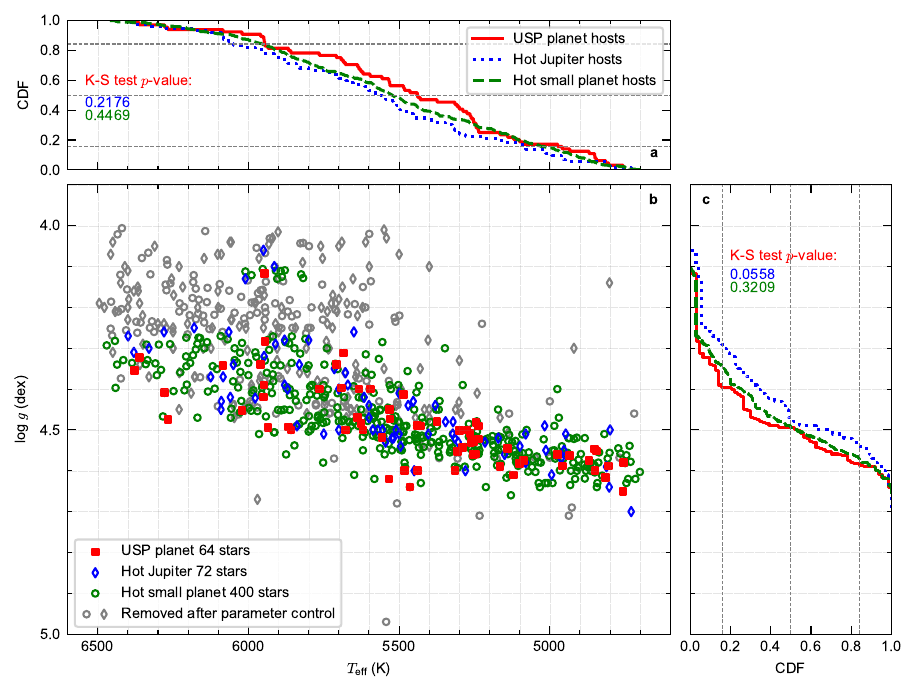}}
{\textbf{\hypertarget{figure_s4}{Supplementary Fig. 4: Comparison of $T_{\rm{eff}}$ and $\log g$ of hot planet hosts, similar to \hyperlink{figure_s3}{Supplementary Fig. 3} but after applying the stellar parameter control in $T_{\rm eff}$ - $\log g$ plane.}} 
USP planet hosts have comparable $T_{\rm{eff}}$ and $\log g$ distributions to hot Jupiter and hot small planet hosts in light of typical uncertainties ($\sim96$ K for $T_{\rm{eff}}$ and $\sim0.04$ dex for $\log g$), with median values of $T_{\rm{eff}}$ at 5439 K, 5562 K, and 5523 K, and median values of $\log g$ at 4.50 dex, 4.48 dex, and 4.50 dex, respectively.
All the K-S test $p-$values of the controlled parameters increased to greater than 0.05 after parameter control.}
\end{figure*}

\hypertarget{Metallicity}{\subsection*{3. Metallicity distribution of hot planet hosts}}
\phantomsection\label{sec:Metallicity}

In this section, we compare the metallicity ($\rm [Fe/H]$) distributions of the host stars in these three planetary populations.

\hyperlink{figure_s5}{Supplementary Fig. 5a} compares [Fe/H] distributions for the sample before parameter control.
The $\rm [Fe/H]$ distribution of USP planet hosts differs from that of HJ hosts by about $2 \sigma$, with a K-S test $p-$value of 0.0611. 
In contrast, the $\rm [Fe/H]$ distribution of USP planet hosts is similar to that of HS hosts, with a K-S test $p-$value of 0.1881.

To minimize the potential impact of differences in $T_{\rm{eff}}$ and $\log g$ in the sample (\hyperlink{figure_s3}{Supplementary Fig. 3}), we apply stellar parameter control in the $T_{\rm{eff}}$-$\log g$ plane, as demonstrated in \hyperlink{Stellar_parameter_control}{Methods \S2} (\hyperlink{figure_s4}{Supplementary Fig. 4}). 
We again compare the $\rm [Fe/H]$ distributions of the host stars of the three planetary populations after parameter control in \hyperlink{figure_s5}{Supplementary Fig. 5b}. 
The K-S test $p-$value result of 0.0590 shows that the $\rm [Fe/H]$ distribution of USP planet hosts is different from that of HJ hosts, corresponding to a difference of about $2 \sigma$. 
However, the $\rm [Fe/H]$ distribution of USP planet hosts is similar to that of HS hosts, as evidenced by a K-S test $p-$value of 0.2613.

Our results show that the $\rm [Fe/H]$ distribution of USP planet hosts is more similar to that of HS hosts than to HJ hosts. 
These results indicate that USP planets are more likely to have a potential common origin with HS planets rather than HJs, aligning with the previous findings of Winn et al. (2017) \citep{2017AJ....154...60W}.

\begin{figure*}[!t]
\centerline{\includegraphics[width=\textwidth]{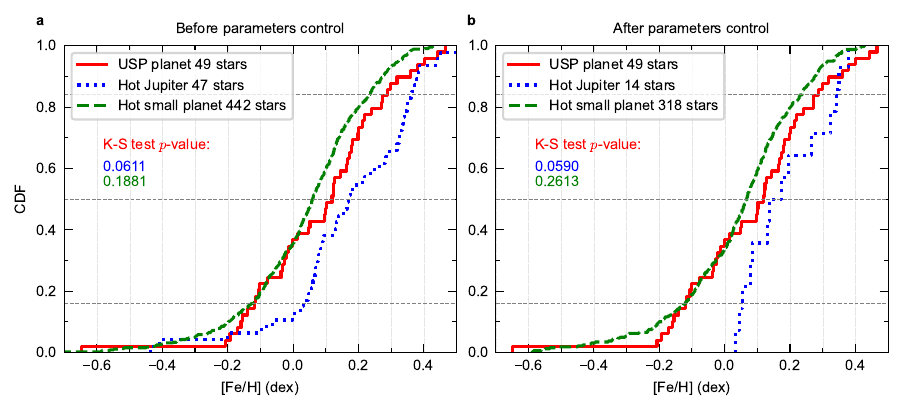}}
{\textbf{\hypertarget{figure_s5}{Supplementary Fig. 5: Comparison of $\rm [Fe/H]$ of hot planet hosts.}} 
Panels show the cumulative distributions of $\rm [Fe/H]$ for the USP planet hosts (red solid lines), hot Jupiter hosts (blue dotted lines), and hot small planet hosts (green dashed lines). 
The horizontal gray dashed lines represent the 50$\pm$34.1 percentiles in the distribution. 
Each panel shows the K-S test $p-$values comparing hot Jupiter and hot small planet hosts to USP planet hosts.
\textbf{a}, The sample before parameter control. 
\textbf{b}, The sample after parameter control in $T_{\rm eff}$ - $\log g$ plane (\hyperlink{figure_s4}{Supplementary Fig. 4}). 
The $\rm [Fe/H]$ of USP planet hosts is more similar to that of hot small planet hosts, rather than hot Jupiter hosts.}
\label{figure_s5}
\end{figure*}

\hypertarget{Kinematic}{\subsection*{4. Kinematic properties of hot planet hosts}}
\phantomsection\label{sec:Kinematic}

\hypertarget{4.1}{\subsubsection*{4.1 Comparison of total velocity and TD/D}}
\phantomsection\label{sec:4.1}

We first compare the distributions of the total velocity ($V_{\rm tot}$) and the relative probabilities between thick disk to thin disk ($TD/D$) of the host stars in these three planetary populations. $V_{\rm tot}$ is defined as:

\begin{equation}
V_{\rm tot} = (U_{\rm LSR}^2+V_{\rm LSR}^2+W_{\rm LSR}^2)^{1/2},
\label{Eq_S4}
\end{equation}
where $U_{\rm LSR}$, $V_{\rm LSR}$, $W_{\rm LSR}$, and $TD/D$ are calculated in \hyperlink{1.3}{Methods \S1.3}. 

\hyperlink{figure_s6}{Supplementary Fig. 6a} and \hyperlink{figure_s7}{Supplementary Fig. 7a} show $V_{\rm tot}$ and $TD/D$ for the sample before parameter control, respectively. 
As shown in the figures, USP planet hosts exhibit higher values of $V_{\rm tot}$ and $TD/D$ in comparison to HJ/HS planet hosts.
Assuming a threshold of $TD/D>2$, about $9.4^{+5.6}_{-3.7}\%$ $(6/64)$ of USP planet hosts, $4.2^{+1.1}_{-0.8}\%$ $(23/550)$ of HS planet hosts, and $4.0^{+1.8}_{-1.3}\%$ $(9/225)$ of HJ hosts are likely to be thick disk stars.
To quantify the statistical significance of these differences, we perform two-sample K-S tests. 
For the ($V_{\rm tot}$) distribution, the $p-$value of the K-S test results for HJ hosts and HS planet hosts versus USP planet hosts are 0.0015 and 0.0384.
Likewise, for the $TD/D$ distribution, the K-S test $p-$values are 0.0056 and 0.0378.

After applying stellar parameter control to the sample (as in \hyperlink{figure_s4}{Supplementary Fig. 4}), we again compare the distributions of $V_{\rm tot}$ and $TD/D$ between different planetary hosts (\hyperlink{figure_1}{Fig. 1} in the main text, corresponding to \hyperlink{figure_s6}{Supplementary Fig. 6b} and \hyperlink{figure_s7}{Supplementary Fig. 7b}). 
Taking a threshold of $TD/D>2$, about $9.4^{+5.6}_{-3.7}\%$ $(6/64)$ of USP planet hosts, $3.8^{+1.2}_{-1.0}\%$ $(15/400)$ of HS planet hosts, and $1.4^{+3.2}_{-1.1}\%$ $(1/72)$ of HJ hosts are likely to be thick disk stars.
For the $V_{\rm tot}$ distribution, the K-S test $p-$values for the comparisons between USP planet hosts and HJ hosts, as well as USP planet hosts and HS planet hosts, are 0.0025 and 0.0359, respectively. 
Similarly, the K-S test for comparing the $TD/D$ distribution between USP-HJ and USP-HS results in $p-$values of 0.0014 and 0.0379, respectively. 
Our results show that USP planet hosts are dynamically hotter than HJ and HS hosts and have a larger fraction of thick disk stars. 

\begin{figure*}[!t]
\centerline{\includegraphics[width=\textwidth]{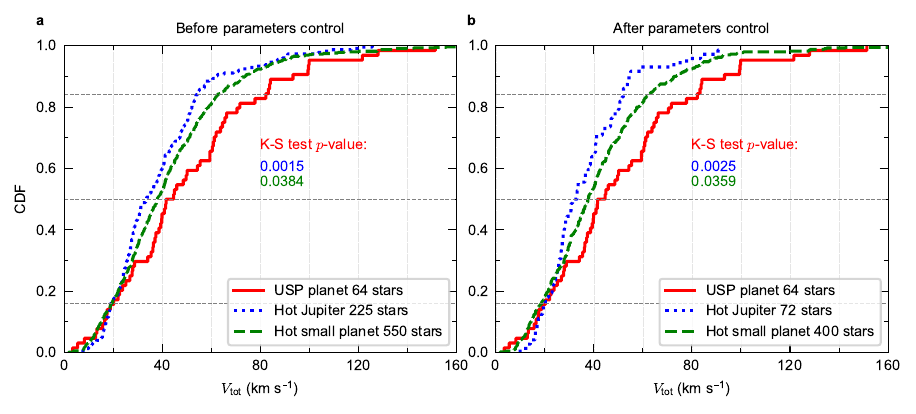}}
{\textbf{\hypertarget{figure_s6}{Supplementary Fig. 6: Similar to \hyperlink{figure_s5}{Supplementary Fig. 5}, but comparison of the total velocities ($V_{\rm tot}$) of hot planet hosts.}} 
USP planet hosts exhibit higher values of $V_{\rm tot}$.}
\end{figure*}

\begin{figure*}[!t]
\centerline{\includegraphics[width=\textwidth]{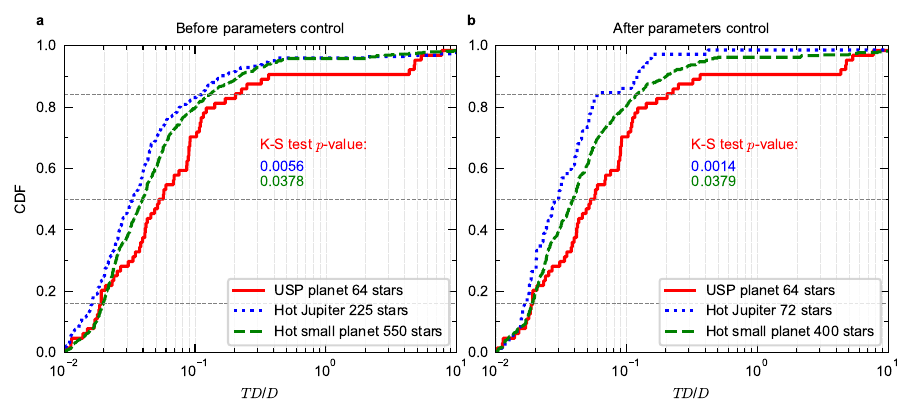}}
{\textbf{\hypertarget{figure_s7}{Supplementary Fig. 7: Similar to \hyperlink{figure_s5}{Supplementary Fig. 5}, but comparison of the relative probabilities between thick disk to thin disk ($TD/D$) of hot planet hosts.}} 
USP planet hosts exhibit higher values of $TD/D$.}
\end{figure*}

\hypertarget{4.2}{\subsubsection*{4.2 Comparison of velocity dispersion distribution}}
\phantomsection\label{sec:4.2}

We further compare the total velocity dispersions of the host stars of these three planetary populations for the sample before and after parameter control.
Following PAST \uppercase\expandafter{\romannumeral1} \citep{2021ApJ...909..115C}, we calculate the total velocity dispersion ($\sigma_{\rm{tot}}$) for host stars of each population as follows:

\begin{equation}
\sigma_{\rm{tot}} = (\sigma_{\rm{U}}^{2}+\sigma_{\rm{V}}^{2}+\sigma_{\rm{W}}^{2})^{\frac{1}{2}},
\label{Eq_S5}
\end{equation}
where $\sigma_{\rm{U}}$, $\sigma_{\rm{V}}$, and $\sigma_{\rm{W}}$ are the velocity dispersions in each direction. 
Their values and uncertainties are obtained by resampling from a normal distribution $N(V,\text{err}_V)$ given the velocity values and uncertainties for each star in different directions. 
The values and uncertainties in $\sigma_{\rm{tot}}$ are set as the 50$\pm$34.1 percentiles in the resampled $\sigma_{\rm{tot}}$ distributions. 

For the sample before parameter control, the Toomre diagrams in \hyperlink{figure_s9}{Supplementary Fig. 9a} show the different distributions of the three planetary hosts in velocity space. As can be seen, USP planet hosts are more spread out in the velocity space, with a $\sigma_{\rm{tot}}$ of $55.27^{+0.17}_{-0.17} \, \rm{km} \, \rm{s}^{-1}$, significantly larger than those of HJ ($42.68^{+0.10}_{-0.10} \, \rm{km} \, \rm{s}^{-1}$) and HS planet ($47.24^{+0.07}_{-0.07} \, \rm{km} \, \rm{s}^{-1}$) hosts. 

In the sample after parameter control, it also shows that USP planet hosts have a larger $\sigma_{\rm{tot}}$ compared to HJ ($39.35^{+0.15}_{-0.15} \, \rm{km} \, \rm{s}^{-1}$) and HS planet ($46.83^{+0.06}_{-0.06} \, \rm{km} \, \rm{s}^{-1}$) hosts. The results are shown in \hyperlink{figure_s8}{Supplementary Fig. 8b} and \hyperlink{figure_s9}{Supplementary Fig. 9b} (corresponding to \hyperlink{figure_2}{Fig. 2} in the main text). 
Our results strongly indicate that USP planet hosts have greater velocity dispersion and are dynamically hotter than HJ and HS hosts. 

\hypertarget{4.3}{\subsubsection*{4.3 Comparison of kinematic age distribution}}
\phantomsection\label{sec:4.3}

The kinematic properties ($V_{\rm tot}$, $TD/D$, and $\sigma_{\rm{tot}}$) are proxies of relative age \citep{2021ApJ...909..115C}.
The above results consistently show the differences in the kinematic properties among these three categories of planetary host stars, indicating that USP planet hosts are likely an older population. 

To demonstrate this, we infer the age distributions of the three planetary populations' hosts from their kinematics by applying the Age-Velocity dispersion Relation (AVR) from PAST \uppercase\expandafter{\romannumeral1} \citep{2021ApJ...909..115C}, which gives

\begin{equation}
t = \left(\frac{\sigma}{k\rm \, km \ s^{-1}}\right)^{\frac{1}{b}}\, \rm Gyr,
\label{Eq_S6}
\end{equation}
where $t$ is stellar age, $\sigma$ is the velocity dispersion, and $k$ and $b$ are two coefficients for AVR. 
We use the total velocity dispersion ($\sigma_{\rm tot}$) calculated in the previous section and the AVR coefficients from \hyperlink{table_s2}{Supplementary Table 2} for the calculations. 
We obtain the kinematic age distribution of the host stars by resampling the values and uncertainties of the AVR coefficients ($k$ and $b$) and the total velocity dispersion ($\sigma_{\rm tot}$). 
The values and uncertainties of the kinematic ages are set as the 50$\pm$34.1 percentiles in the resampled age distribution. 
The solid black line in \hyperlink{figure_s8}{Supplementary Fig. 8} shows the AVR we adopted.

In the Toomre diagrams, we show the velocity space distribution and kinematic ages of the sample before parameter control (\hyperlink{figure_s9}{Supplementary Fig. 9a}). 
The kinematic ages of the USP planet hosts, HS planet hosts, and HJ hosts are $5.71^{+0.72}_{-0.58}$ Gyr, $3.85^{+0.42}_{-0.35}$ Gyr, and $2.98^{+0.30}_{-0.25}$ Gyr, respectively. 
Furthermore, out of 10,000 Monte Carlo resampling cases, the calculated kinematic ages of USP planet host stars exceed those of HJ hosts in all 10,000 cases, and they are also older than the ages of HS planet hosts in 9,969 cases. 
This translates to a confidence level of over 99.99\% for the statement that USP planet hosts are older than HJ hosts and 99.69\% for the statement that USP planet hosts are older than HS planet hosts.
As can be seen, USP planet hosts are significantly older (also see \hyperlink{figure_s8}{Supplementary Fig. 8a} for comparison).

The results after parameter control are shown in \hyperlink{figure_s8}{Supplementary Fig. 8b} and \hyperlink{figure_s9}{Supplementary Fig. 9b} (corresponding to \hyperlink{figure_2}{Fig. 2} in the main text). 
The derived ages are $5.71^{+0.72}_{-0.58}$ Gyr for USP planet hosts, $3.77^{+0.41}_{-0.34}$ Gyr for HS planet hosts, and $2.44^{+0.23}_{-0.19}$ Gyr for HJ hosts.
Out of the 10,000 sets of resampled data, the kinematic ages of USP planet hosts are older than HJ hosts in 10,000 trials, corresponding to a confidence level of over 99.99\%, and older than HS planets in 9,978 trials, corresponding to a confidence level of 99.78\%. 
In summary of this section, our results indicate that the USP planet hosts are dynamically hotter and older than the other two populations.

\begin{figure*}[!t]
\centerline{\includegraphics[width=\textwidth]{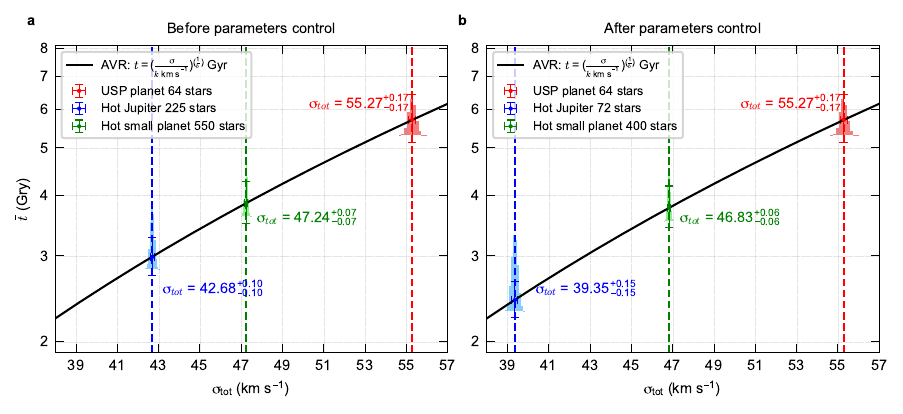}}
{\textbf{\hypertarget{figure_s8}{Supplementary Fig. 8: Comparison of total velocity dispersions ($\sigma_{\rm{tot}}$) and the kinematic ages.}}
The USP planet hosts are colored in red, hot Jupiter hosts in blue, and hot small planet hosts in green.
The colored histograms represent the distributions of velocity dispersions obtained from 10,000 resamplings.
The colored vertical dashed lines indicate the medians of the velocity dispersion distributions.
We derive the kinematic ages using the Age-Velocity dispersion Relation (AVR, Equation 7 in the Methods) represented by the black solid line.
The data points with error bars show the median values and $\pm1\sigma$ ranges of $\sigma_{\rm{tot}}$ and the kinematic ages.
USP planet hosts have larger $\sigma_{\rm{tot}}$ and are dynamically older. 
\textbf{a}, The sample before parameter control. 
\textbf{b}, The sample after parameter control in $T_{\rm eff}$ - $\log g$ plane (\hyperlink{figure_s4}{Supplementary Fig. 4}). 
USP planet hosts have larger $\sigma_{\rm{tot}}$ and are dynamically older.}
\label{figure_s8}
\end{figure*}

\begin{figure*}[!t]
\centerline{\includegraphics[width=\textwidth]{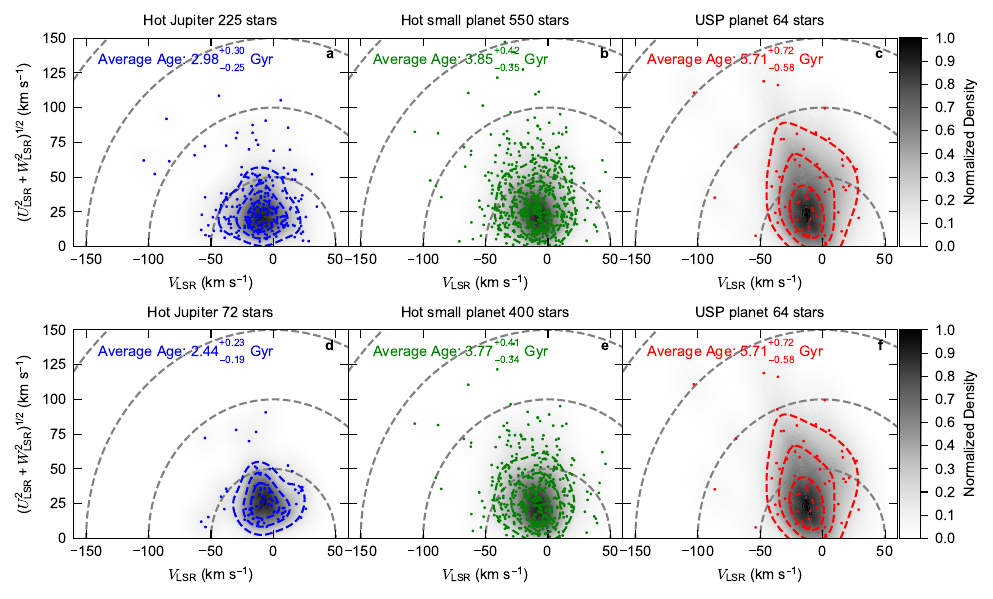}}
{\textbf{\hypertarget{figure_s9}{Supplementary Fig. 9: The Toomre diagrams.}} 
The USP planet hosts are colored in red, hot Jupiter hosts in blue, and hot small planet hosts in green.
The gray dashed lines show constant values of the total Galactic velocity $V_{\rm tot}$ in steps of 50 $\rm{km} \, \rm{s}^{-1}$.
We plot the normalized number density contour lines, representing values of 0.25, 0.5, and 0.75.
In each panel, we print the kinematic ages of the corresponding host stars.
\textbf{a-c}, the sample before parameter control. 
\textbf{d-e}, the sample after parameter control in $T_{\rm eff}$ - $\log g$ plane (\hyperlink{figure_s4}{Supplementary Fig. 4}).
USP planet hosts are more spread out in the velocity space and dynamically older.}
\end{figure*}

\begin{table}[!t]
\centering
\renewcommand\arraystretch{1.5}
\hypertarget{table_s2}{{\bfseries Supplementary Table 2: fitting parameters of the Age-Velocity dispersion relation (AVR) from PAST \uppercase\expandafter{\romannumeral1} \citep{2021ApJ...909..115C}.}}
{\footnotesize
\begin{tabular}{l|cccccc} \hline
                & \multicolumn{2}{c}{---------~~$k \ \rm (km \ s^{-1})$~~---------} & \multicolumn{2}{c}{---------~~$b$~~---------}    \\ 
               &  value &  $1-\sigma$ interval &  value &  $1-\sigma$ interval  \\ \hline
    $U$  & $23.66$ & (23.07, 24.32) & $0.34$ & $(0.33,0.36)$  \\
    $V$  & $12.49$ & (12.05, 12.98) & $0.43$ & $(0.41,0.45)$  \\
    $W$  & $8.50$ & (8.09, 8.97) & $0.54$ & $(0.52,0.56)$ \\
    $V_{\rm tot}$ & $27.55$ & (26.84, 28.37) & $0.40$ & $(0.38, 0.42)$ \\ \hline
\end{tabular}}
\end{table}

\hypertarget{Frequency}{\subsection*{5. Frequency of USP planets as a function of age}}
\phantomsection\label{sec:Frequency}

To explore the evolutionary process of the occurrence rate of USP planets among hot small mass planets with increasing age, we divide the samples of USP planets and HS planets and their hosts from \hyperlink{Data_sample}{Methods \S1} into three different subsamples according to their hosts' relative probabilities between thick disk to thin disk ($TD/D$). 
As $TD/D$ serves as a proxy for relative age \citep{2021ApJ...909..115C}, the three subsamples represent younger, intermediate, and older hot small mass planet systems, respectively. 
Then we calculate the relative frequency of USP planets among all hot small mass planets ($f_{\rm USP}$) with the following formula: 

\begin{equation}
f_{\rm USP} = \frac{N_{\rm USP}}{N_{\rm USP}+N_{\rm HS}},
\label{Eq_S7}
\end{equation}
where $N_{\rm USP}$ and $N_{\rm HS}$ are the number of USP and HS planets, respectively. 

According to the AVR, the kinematic ages of hosts in the subsamples are $0.67^{+0.05}_{-0.05}$ Gyr, $2.17^{+0.19}_{-0.16}$ Gyr, and $9.89^{+1.55}_{-1.18}$ Gyr, respectively.
The corresponding $f_{\rm USP}$ values from the younger to the older subsample are $0.0846^{+0.0259}_{-0.0203}$ $(17/201)$, $0.0842^{+0.0258}_{-0.0202}$ $(17/202)$, and $0.1493^{+0.0325}_{-0.0271}$ $(30/201)$, respectively. 

To investigate the age-$f_{\rm USP}$ relation for the kinematic sample, we fit a simple linear model:

\begin{equation}
f_{\rm USP} = A \times \log_{10}(t/\rm Gyr) + B,
\label{Eq_S8}
\end{equation}
where $A$ and $B$ are the parameters determined through fitting. Given the data and uncertainties, we resample and refit the data points 10,000 times to obtain the best-fitting parameters and their error bars. The resulting best-fit model is: 

\begin{equation}
f_{\rm USP} = 0.571^{+0.0328}_{-0.0324} \times \log_{10}(t/\rm Gyr) + 0.0843^{+0.0168}_{-0.0173} \ (\rm{before}\ \rm{parameters}\ \rm{control}).
\label{Eq_S9}
\end{equation}

Out of 10,000 resampling and refitting iterations, $f_{\rm USP}$ positively correlates with age in 9,641 cases, corresponding to a confidence level of 96.41\%. 
These results indicate that $f_{\rm USP}$ is highest in the older subsample and appears to be positively correlated with age, as shown in \hyperlink{figure_s1}{Supplementary Fig. 1}. 

However, subsamples from the same source exhibit differences in other stellar parameters (e.g., $\rm [Fe/H]$ and $T_{\rm{eff}}$, see \hyperlink{figure_s11}{Supplementary Fig. 11a,e}).
To isolate the effect of stellar age, we select the intermediate subsamples as the standard sample and apply the stellar parameter control method from \hyperlink{Stellar_parameter_control}{Methods \S2} in the $T_{\rm{eff}}$-$\rm [Fe/H]$ plane ($w_{2}=0$). 
Notably, we exclude stars without $\rm [Fe/H]$ measurements during the parameter control.

After controlling for stellar parameters (see \hyperlink{figure_s12}{Supplementary Fig. 12a,e}), we recalculate the average age and $f_{\rm USP}$ for each subsample. 
As shown in \hyperlink{figure_s10}{Supplementary Fig. 10b} (as in \hyperlink{figure_3}{Fig. 3}), for the kinematic sample, the kinematic ages of the younger, intermediate, and older subsamples are $0.55_{-0.04}^{+0.04}$ Gyr, $2.00_{-0.14}^{+0.17}$ Gyr, and $9.46_{-1.12}^{+1.46}$ Gyr, respectively. 
We find that $f_{\rm USP}$ increases with age, with values of $0.0538^{+0.0364}_{-0.0232}$ $(5/93)$, $0.0877^{+0.0290}_{-0.0224}$ $(15/171)$, and $0.1546^{+0.0511}_{-0.0395}$ $(15/97)$ for the younger, intermediate, and older subsamples, respectively.
The best fit for the kinematic sample is: 

\begin{equation}
f_{\rm USP} = 0.0821^{+0.0446}_{-0.0438} \times \log_{10}(t/\rm Gyr) + 0.0713^{+0.0197}_{-0.0202} \ (\rm{after}\ \rm{parameters}\ \rm{control}),
\label{Eq_S10}
\end{equation}
and $f_{\rm USP}$ is positively correlated with age in 9,690 out of 10,000 refitting sets, indicating a confidence level of 96.90\%, which is slightly higher than the result before parameter control. 

As a consistency check, we also calculate $f_{\rm USP}$ as a function of age by adopting other age estimates from other methods and sources as follows:

\begin{enumerate}
\item Berger2023: Isochrone ages from Berger et al. (2023) \citep{2023arXiv230111338B} with a median uncertainty of $\sim85\%$. 
Berger et al. (2023) obtained a homogeneous catalog of {\it Kepler}, K2, and TESS planets and their host stars using Gaia spectrophotometric metallicities and CKS metallicities. 
We keep the Sun-like stars with relatively reliable isochrone ages by applying the selection criteria, i.e., metallicity provenance with {\it Poly}, isochrone age $<$ 12 Gyr. 
We crossmatch our sample with Berger et al. (2023) to obtain the isochrone ages. 
The selected sample contains 42 USP planets and 459 HS planets. The median age uncertainty in the selected sample is $\sim105\%$.
\item NASAarchive: The NASA exoplanet archive \citep{2013PASP..125..989A}. 
We select Sun-like stars with ages smaller than 12 Gyr hosting USP and HS planets. 
The selected sample contains 60 USP planets and 765 HS planets. The median age uncertainty in the selected sample is $\sim64\%$. 
\item EU: The Extrasolar Planets Encyclopaedia (\url{https://exoplanet.eu}). 
We also select Sun-like stars with ages smaller than 12 Gyr hosting USP and HS planets. 
The selected sample contains 54 USP planets and 508 HS planets. The median age uncertainty in the selected sample is $\sim55\%$. 
\end{enumerate}

We divide the samples with ages obtained from other methods and sources into different subsamples based on age, then calculate the average age and $f_{\rm USP}$ for each subsample. 
Similarly, we use the intermediate subsamples as the standard sample for each source and apply parameter control in the $T_{\rm{eff}}$-$\rm [Fe/H]$ plane.
As shown in \hyperlink{figure_s10}{Supplementary Fig. 10}, the results obtained using other methods to estimate age are aligned with those derived from kinematic age. 
In summary of this section, our results indicate that the frequency of USP planets ($f_{\rm USP}$) increases with age.

\begin{figure*}[!t]
\centerline{\includegraphics[width=\textwidth]{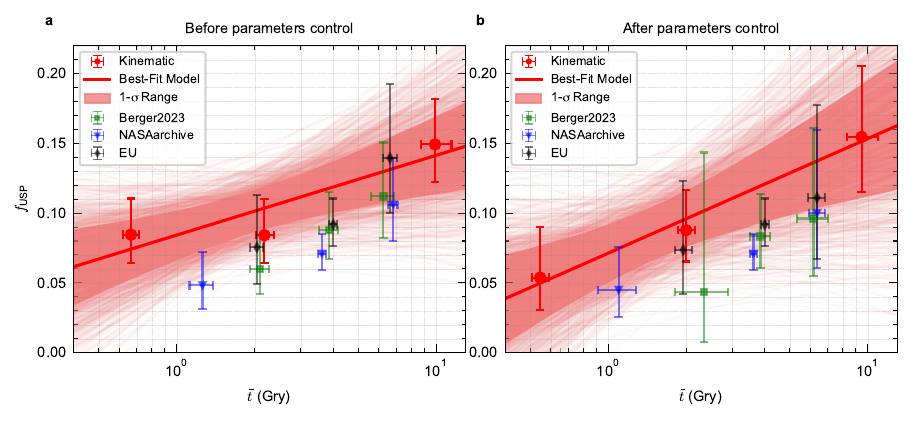}}
{\textbf{\hypertarget{figure_s10}{Supplementary Fig. 10: The frequencies of USP planets ($f_{\rm{USP}}$) as functions of average age from different indicators.}} 
The data points with error bars show the median values and $\pm1\sigma$ ranges of kinematic ages on the horizontal axis, and the frequencies with Poisson counting errors on the vertical axis.
(1) Kinematic (red dots): The kinematic ages from the Age-Velocity dispersion Relation (AVR) in PAST \uppercase\expandafter{\romannumeral1} and \uppercase\expandafter{\romannumeral2} \citep{2021ApJ...909..115C,2022AJ....163..249C}. 
(2) Berger2023 (green squares): Isochrone ages from Berger et al. 2023 \citep{2023arXiv230111338B}. 
(3) NASAarchive (blue inverted triangles): The ages from the NASA exoplanet archive \citep[https://exoplanetarchive.ipac.caltech.edu;][]{2013PASP..125..989A}. 
(4) EU (black diamonds): The ages from the Extrasolar Planets Encyclopaedia (http://exoplanet.eu/). 
The red solid line and shaded region denote the best-fit linear model and 1 - $\sigma$ interval for the kinematic sample. 
The thin line of light red represents the results of 10,000 fits from resampling the kinematic sample. 
\textbf{a}, The sample before parameter control.
The best-fit linear model follows $f_{\rm USP} = 0.571^{+0.0328}_{-0.0324} \times \log_{10}(t/\rm Gyr) + 0.0843^{+0.0168}_{-0.0173}$ (Equation 10 in Methods).
\textbf{b}, The sample after parameter control in $T_{\rm{eff}}$-$\rm [Fe/H]$ plane (\hyperlink{figure_s12}{Supplementary Fig. 12}). 
The best-fit linear model follows $f_{\rm USP} = 0.0821^{+0.0446}_{-0.0438} \times \log_{10}(t/\rm Gyr) + 0.0713^{+0.0197}_{-0.0202}$ (Equation 11 in Methods).}
\end{figure*}

\begin{figure*}[!t]
\centerline{\includegraphics[width=\textwidth]{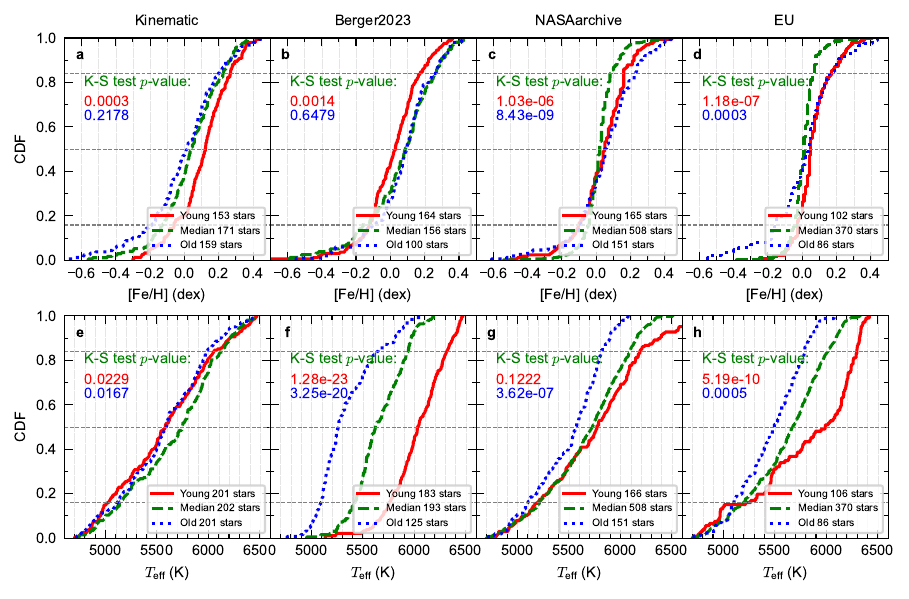}}
{\textbf{\hypertarget{figure_s11}{Supplementary Fig. 11: Comparison of $\rm [Fe/H]$ and $T_{\rm{eff}}$ for stellar samples of different ages before parameter control.}} 
Panels show the cumulative distributions of $\rm [Fe/H]$ (\textbf{a-d}) and $T_{\rm{eff}}$ (\textbf{e-h}) for the younger (red solid lines), intermediate (green dashed lines), and older (blue dotted lines) subsamples in \hyperlink{Frequency}{Methods \S5}.
The horizontal gray dashed lines represent the 50$\pm$34.1 percentiles in the distribution. 
The $\rm [Fe/H]$ and $T_{\rm{eff}}$ distributions in our kinematic sample for intermediate subsample exhibit significant differences with those of younger and older subsample in light of typical uncertainties ($\sim0.05$ dex for $\rm [Fe/H]$ and $\sim96$ K for $T_{\rm{eff}}$), with median values of $\rm [Fe/H]$ at 0.04 dex, 0.12 dex, and 0.01 dex, and median values of $T_{\rm{eff}}$ at 5747 K, 5582 K, and 5591 K, respectively.
Each panel shows the K-S test $p-$values comparing the younger and older subsamples to the intermediate subsample.}
\end{figure*}

\begin{figure*}[!t]
\centerline{\includegraphics[width=\textwidth]{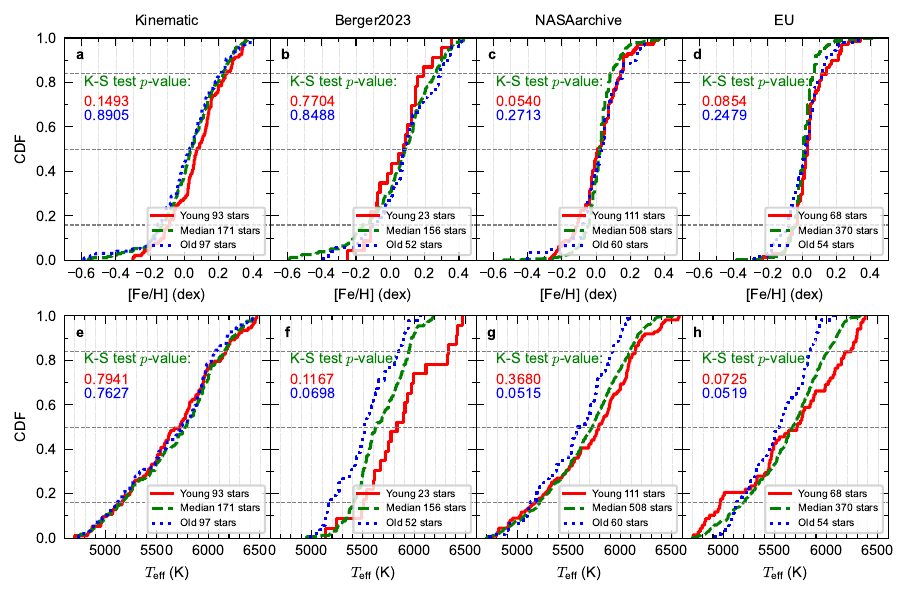}}
{\textbf{\hypertarget{figure_s12}{Supplementary Fig. 12: Similar to \hyperlink{figure_s11}{Supplementary Fig. 11}, but after applying the stellar parameter control in $\rm [Fe/H]$ and $T_{\rm{eff}}$ plane.}}
The intermediate subsample in our kinematic sample has comparable $\rm [Fe/H]$ and $T_{\rm{eff}}$ distributions to younger and older subsample in light of typical uncertainties ($\sim0.05$ dex for $\rm [Fe/H]$ and $\sim96$ K for $T_{\rm{eff}}$), with median values of $\rm [Fe/H]$ at 0.04 dex, 0.08 dex, and 0.03 dex, and median values of $T_{\rm{eff}}$ at 5777 K, 5723 K, and 5746 K, respectively.
All the K-S test $p-$values of the controlled parameters increased to greater than 0.05 after parameter control.} 
\end{figure*}

\hypertarget{architecture}{\subsection*{6. Age dependence of the architecture of USP planet systems}}
\phantomsection\label{sec:architecture}

In this section, we investigate the age-dependent variations in the architecture of USP planetary systems. 
Specifically, we compare the orbital period, orbital spacing, and transiting multiplicity of younger and older hot planetary systems. 

We divide the samples of USP planets and HS planets and their hosts from \hyperlink{Data_sample}{Methods \S1} into two subsamples according to their hosts' relative probabilities between thick disk to thin disk ($TD/D$): a younger one and an older one. 
The younger subsample contains 40\% ($N_{\rm {star}} = 242$) of the planetary systems, while the older subsample contains 60\% ($N_{\rm {star}} = 362$) of the planetary systems, with corresponding kinematic ages of $0.79^{+0.06}_{-0.05}$ Gyr and $6.46^{+0.86}_{-0.69}$ Gyr, respectively. 
\hyperlink{figure_s15}{Supplementary Fig. 15} shows an overview of planetary architecture in the younger and older subsamples. 

To isolate the age effect, we use the younger subsample as the standard sample and apply parameter control to the older subsample in the $T_{\rm{eff}}$-$\rm [Fe/H]$ plane ($w_{2}=0$). 
Specifically, we limit the maximum value of the weighted Euclidean distance ($d$) of host stars in the $T_{\rm{eff}}$-$\rm [Fe/H]$ plane to ensure that the number of stars in the older subsample after parameter control matches that of the younger subsample after parameter control. 
\hyperlink{figure_s13}{Supplementary Fig. 13} shows the $T_{\rm{eff}}$ and $\rm [Fe/H]$ distributions of the subsamples before and after control, as well as the $p-$value results of the two-sample K-S tests. 
Notably, we exclude stars without $\rm [Fe/H]$ measurements during the parameter control.

After parameter control, the kinematic ages of the younger and older subsamples are $0.69_{-0.05}^{+0.05}$ Gyr and $5.53_{-0.56}^{+0.70}$ Gyr, respectively.
An overview of the planetary architecture of the younger and older subsamples after parameter control is shown in \hyperlink{figure_s16}{Supplementary Fig. 16}.

\begin{figure*}[!t]
\centerline{\includegraphics[width=\textwidth]{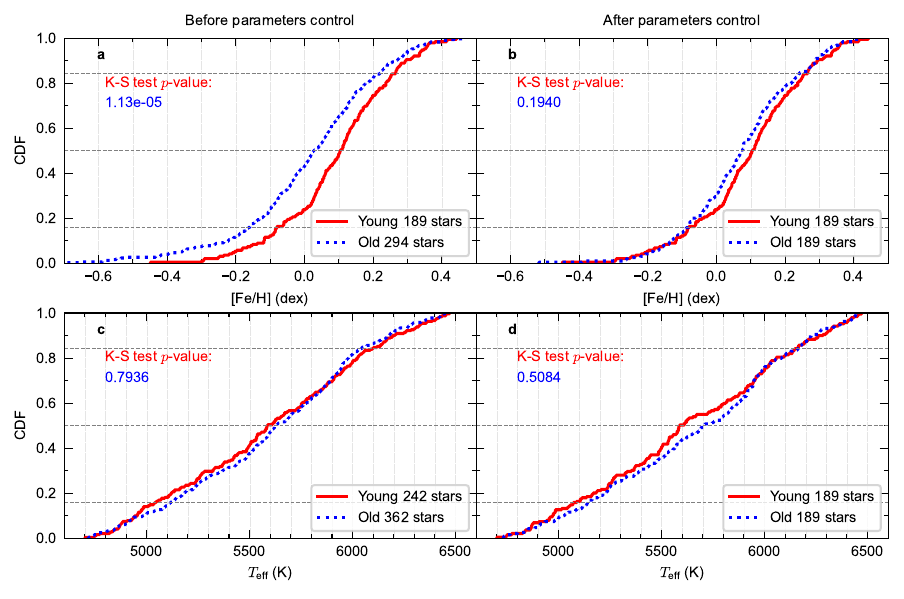}}
{\textbf{\hypertarget{figure_s13}{Supplementary Fig. 13: Comparison of $\rm [Fe/H]$ and $T_{\rm{eff}}$ for stellar samples of different ages.}} 
Panels show the cumulative distributions of $\rm [Fe/H]$ (\textbf{a-b}) and $T_{\rm{eff}}$ (\textbf{c-d}) for the younger (red solid lines) and older (blue dotted lines) subsamples in \hyperlink{architecture}{Methods \S6}.
The horizontal gray dashed lines represent the 50$\pm$34.1 percentiles in the distribution.
\textbf{a and c}, the sample before parameter control.
The $\rm [Fe/H]$ and $T_{\rm{eff}}$ distributions for younger subsample exhibit significant differences with those of older subsample in light of typical uncertainties ($\sim0.05$ dex for $\rm [Fe/H]$ and $\sim96$ K for $T_{\rm{eff}}$), with median values of $\rm [Fe/H]$ at 0.10 dex and 0.03 dex, and median values of $T_{\rm{eff}}$ at 5587 K and 5627 K, respectively.
\textbf{b and d}, the sample after parameter control. 
The older subsample has similar $\rm [Fe/H]$ and $T_{\rm{eff}}$ distributions to those of the younger subsample.
The younger subsample has comparable $\rm [Fe/H]$ and $T_{\rm{eff}}$ distributions to the older subsample, with median values of $\rm [Fe/H]$ at 0.10 dex and 0.07 dex, and median values of $T_{\rm{eff}}$ at 5587 K and 5704 K, respectively.
Each panel shows the K-S test $p-$values comparing the older to younger subsample distributions.
After control, all K-S $p-$values exceed 0.15.}
\end{figure*}

\begin{figure*}[!t]
\centerline{\includegraphics[width=\textwidth]{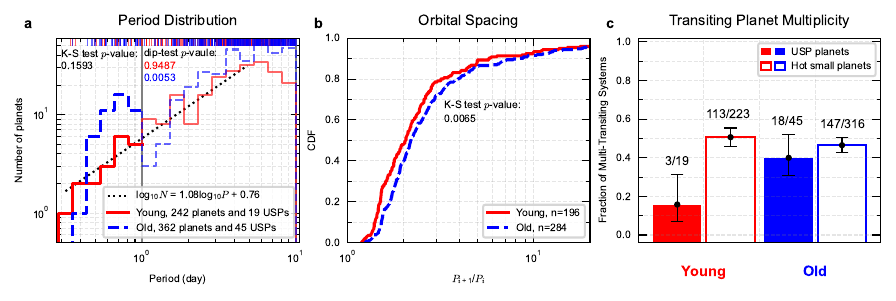}}
{\textbf{\hypertarget{figure_s14}{Supplementary Fig. 14: Similar to \hyperlink{figure_4}{Fig. 4} in the main text, but before applying the stellar parameter control.}}}
\end{figure*}

\hypertarget{6.1}{\subsubsection*{6.1 Orbital period distribution}}
\phantomsection\label{sec:6.1}

We compare the period distribution of the innermost planets in hot planet systems between the younger and older subsamples. 
\hyperlink{figure_s14}{Supplementary Fig. 14a} and \hyperlink{figure_4}{Fig. 4a} show the comparisons of the period distributions before and after parameter control, respectively. 
As can be seen, the period distribution of the younger subsample differs from that of the older subsample.
Before (after) parameter control, the K-S test result for the two period distributions gives a $p-$value of 0.1593 (0.0632).
The period distribution of the younger subsample resembles a power-law distribution, while this is apparently not the case for the older subsample. 
To verify this, we fit the period distribution of the younger subsample within the range of 0-5 days using a simple power-law distribution. The fitted model is: 

\begin{equation}
\log_{10} N = A\times\log_{10} P + B,
\label{Eq_S11}
\end{equation}
where $A$ and $B$ are coefficients to be fitted.

Before parameter control, the K-S test $p-$value for the period distribution of the younger subsample, when compared to the best-fit model, is 0.9436.
This corresponds to a confidence level of 94.36\%, indicating a strong alignment with the power-law distribution. 
In contrast, for the older subsample, the K-S test $p-$value, compared to the best-fit model, is 0.0942, corresponding to a confidence level of $1-0.0942=90.58\%$ that rejects the hypothesis of being drawn from the power-law distribution. 

After applying parameter control, the K-S test $p-$value for the period distribution of the younger subsample compared to the best-fit model is 0.9792, indicating with 97.92\% confidence that it originates from the power-law distribution. 
However, for the older subsample, the K-S test $p-$value, compared to the best-fit model, is 0.1410, suggesting a deviation from the power-law distribution at a $1-0.1410=85.90\%$ confidence level.

Moreover, the difference between the period distribution of the older subsample and the power-law distribution reveals a notable dip/pileup feature around a period of $\sim 1$ day. 
To verify this observation, we perform Hartigan's dip test \citep{10.1214/aos/1176346577} on the period distribution of each subsample in $\log P$ for $P$ in the range 0-2.5 days, to assess the significance of the difference between the period distribution and a unimodal distribution.

Before controlling for stellar parameters, the dip test conducted on the younger subsamples yields a $p-$value of 0.9487, corresponding to a confidence level of 94.87\% that supports the absence of a dip feature. 
Conversely, for the older subsamples, the dip test reveals a $p-$value of 0.0053, which translates to a confidence level of $1-0.0053=99.47\%$ supporting the presence of a dip feature.

After controlling for stellar parameters, the dip/pileup feature remains discernible, albeit with a reduced sample size. 
Specifically, the dip test on the younger subsample produces a $p-$value of 0.9578, indicating a confidence level of 95.78\% that there is no dip feature. 
On the other hand, the dip test for the older subsample returns a $p-$value of 0.0362, corresponding to a confidence level of $1-0.0362=96.38\%$, which suggests the presence of a dip feature. 

In summary of this subsection, our results indicate that the period distribution varies over age and reveals an age-dependent dip/pileup feature in the period distribution of close-in planets.

\hypertarget{6.2}{\subsubsection*{6.2 Orbital period ratio distribution}}
\phantomsection\label{sec:6.2}

We compare the orbital spacing, specifically the orbital period ratio of adjacent planets, among hot multiple planet systems in our sample.
As can be visually seen from \hyperlink{figure_s15}{Supplementary Fig. 15}, the planets are more spread out (i.e., larger orbital spacing) in older subsample (\hyperlink{figure_s15}{Supplementary Fig. 15b}) than in younger subsample (\hyperlink{figure_s15}{Supplementary Fig. 15a}).

This is confirmed by \hyperlink{figure_s14}{Supplementary Fig. 14b}, which compares the cumulative distributions of the period ratios between the younger and older subsamples before parameter control. 
As can be seen, the older subsample exhibits a shift towards a larger period ratio (with a median of 2.18) compared to the younger subsample (with a median of 2.04).
The two-sample K-S test conducted on these subsamples gives a $p-$value of 0.0065, corresponding to the two distributions being distinct with a confidence level of $1-0.0065=99.35\%$.

Furthermore, we divide the period ratios into three groups based on their locations in the multiple planet systems, namely inner pairs ($P_{2}/P_{1}$), middle pairs ($P_{3}/P_{2}$), and outer pairs ($P_{4}/P_{3}$).
As shown in \hyperlink{figure_s17}{Supplementary Fig. 17}, the older subsample exhibits larger period ratios than the younger subsample in all three groups:
inner pairs (medians of 2.37 vs. 2.18), middle pairs (medians of 2.06 vs. 1.86), and outer pairs (medians of 1.83 vs. 1.60). 
The corresponding two-sample K-S test $p-$values are 0.1193, 0.1127, and 0.0912, translating to confidence levels of 88.07\%, 88.73\%, and 90.88\%, respectively. 

These results remain significant after controlling for stellar parameters (\hyperlink{figure_s16}{Supplementary Fig. 16}).
\hyperlink{figure_4}{Fig. 4b} compares the period ratios after parameter control.
Specifically, the older subsample (median of 2.13) maintains a larger period ratio than the younger subsample (median of 1.94). 
The two-sample K-S test returns a $p-$value of 0.0061, confirming the distinctness of the two distributions with a confidence level of $1-0.0061=99.39\%$.
Moreover, the period ratios in the older subsample remain larger than those in the younger subsample across various pairs: 
inner pairs (medians of 2.35 vs. 2.16), middle pairs (medians of 2.05 vs. 1.73), and outer pairs (medians of 1.83 vs. 1.62).
The corresponding two-sample K-S test $p-$values are 0.1048, 0.0176, and 0.2784, which correspond to confidence levels of 89.52\%, 98.24\%, and 72.16\%, respectively (\hyperlink{figure_s18}{Supplementary Fig. 18}).

In summary of this subsection, our results indicate that the older hot multiple planet systems have larger period ratios than the younger ones, and this difference is independent of the planets' locations within the system. 

\hypertarget{6.3}{\subsubsection*{6.3 Transiting planet multiplicity}}
\phantomsection\label{sec:6.3}

We compare the transiting planet multiplicities of the younger and older subsamples of USP and HS planet systems. 
\hyperlink{figure_s14}{Supplementary Fig. 14c} and \hyperlink{figure_4}{Fig. 4c} show the comparisons of the transiting planet multiplicities before and after controlling for parameters, respectively. 

Before controlling for parameters, 113 out of 223 ($\sim50.67^{+4.77}_{-4.77}\%$) HS planet systems in the younger subsample have multiple transiting planets, which is comparable to that of the older subsample ($147/316, \sim46.52^{+3.84}_{-3.84}\%$). 
However, in the case of USP planet systems, the younger subsample ($3/19, \sim15.79^{+15.36}_{-8.59}\%$) has a lower fraction of multiple transiting systems compared to the older subsample ($18/45, \sim40.00^{+11.82}_{-9.34}\%$). 
On the other hand, the fraction of USP planet systems with non-transiting planetary companions is higher in the younger subsample ($3/19, \sim15.79^{+15.36}_{-8.59}\%$) than in the older subsample ($3/45, \sim6.67^{+6.48}_{-3.63}\%$) (see \hyperlink{figure_s15}{Supplementary Fig. 15}). 
When focusing solely on multiple planet systems, we find that, in younger USP planet systems, multiple planet systems are dominated (3 out of 5) by systems \emph{with} non-transiting planets. 
Conversely, in older USP planet systems, multiple planet systems are dominated (16 out of 19) by systems \emph{without} non-transiting planets.

After controlling for stellar parameters, the results are even more significant. 
For HS planet systems, the fraction of multiple transiting systems remains comparable between the younger ($95/176, \sim53.98^{+5.54}_{-5.54}\%$) and older ($87/164, \sim53.05^{+5.69}_{-5.69}\%$) subsamples. 
Conversely, for USP planet systems, the younger subsample ($0/13, \sim0.00^{+14.16}\%$) has a much lower fraction of multiple transiting systems than the older subsample ($10/25, \sim40.00^{+17.06}_{-12.43}\%$). 
Nevertheless, 2 out of the 13 ($\sim15.38^{+20.29}_{-9.94}\%$) USP planet systems in the younger subsample have non-transiting planetary companions, which is higher than the fraction in the older subsample where none are found among 25 USP planet systems (see \hyperlink{figure_s16}{Supplementary Fig. 16}). 
When focusing solely on multiple planet systems, we find that, in younger USP planet systems, multiple planet systems are 100\% (2 out of 2) dominated by systems \emph{with} non-transiting planets. 
Conversely, in older USP planet systems, multiple planet systems are 100\% (10 out of 10) dominated by systems \emph{without} non-transiting planets. 

In summary of this subsection, we find that younger USP planet systems exhibit a lower fraction of multiple transiting systems but a higher fraction of systems with non-transiting planetary companions, compared to older ones.

\begin{figure*}[p]
\centering
\includegraphics[width=0.90\textwidth]{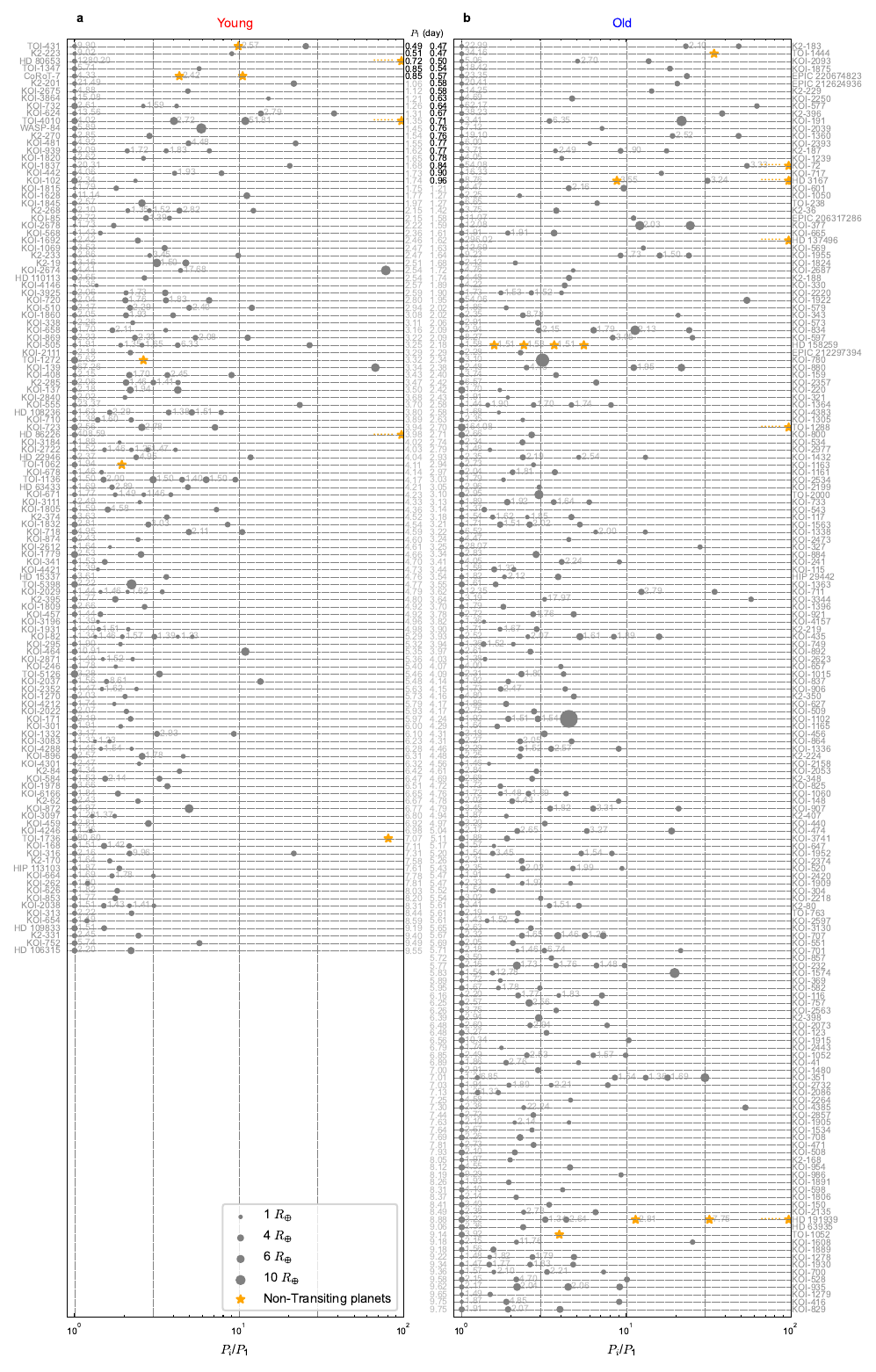}
\caption*{\textbf{\hypertarget{figure_s15}{Supplementary Fig. 15:}} The figure's caption is displayed on the next page.}
\end{figure*}

\clearpage

\begin{figure*}[!t]
\captionof*{figure}{\textbf{Supplementary Fig. 15: Overview of the orbital architectures of planetary systems, categorized into younger (a) and older (b) subsamples before controlling stellar parameters (i.e., $T_{\rm{eff}}$ and $\rm [Fe/H]$).} 
Each dot signifies a planet or planet candidate, and each line connecting the dots depicts a planetary system, with the system's name appearing on both edges of the figure. 
Grey dots represent transiting planets, whereas yellow star symbols denote non-transiting planets. 
The size of the dots, except the yellow ones, scales with the planetary radius.
A number indicates the orbital period ratio between each pair of adjacent planets. 
The periods of the innermost planets within each system are listed in ascending order, with USPs emphasized in black font, from top to bottom in the central area of the figure, located between the two panels. 
The X-axis of each panel is the orbital period normalized by the period of the innermost planets ($P_i/P_{1}$). 
Several long period planets (all of which are non-transiting) with $P_i/P_{1}>100$ are plotted at the right end of each panel and marked with ellipsis on the left. 
\emph{Some patterns emerge: (1) Planets are more spread out (i.e., larger orbital spacing) in older subsample (\textbf{b}) than in younger subsample (\textbf{a}), (see also in \hyperlink{6.2}{Methods \S6.2} and \hyperlink{figure_s17}{Supplementary Fig. 17}). (2) Systems with shorter innermost periods (i.e., $P_1<3-4$ days) tend to have a higher occurrence of non-transiting planets \citep[suggestive of larger mutual inclinations and consistent with observations by Dai et al. 2018,][]{2018ApJ...864L..38D} than systems with longer $P_1$. (3) Younger USP planet systems exhibit a larger fraction of non-transiting planets than older USP planet systems.}}
\end{figure*}

\clearpage

\begin{figure*}[!t]
\centerline{\includegraphics[width=\textwidth]{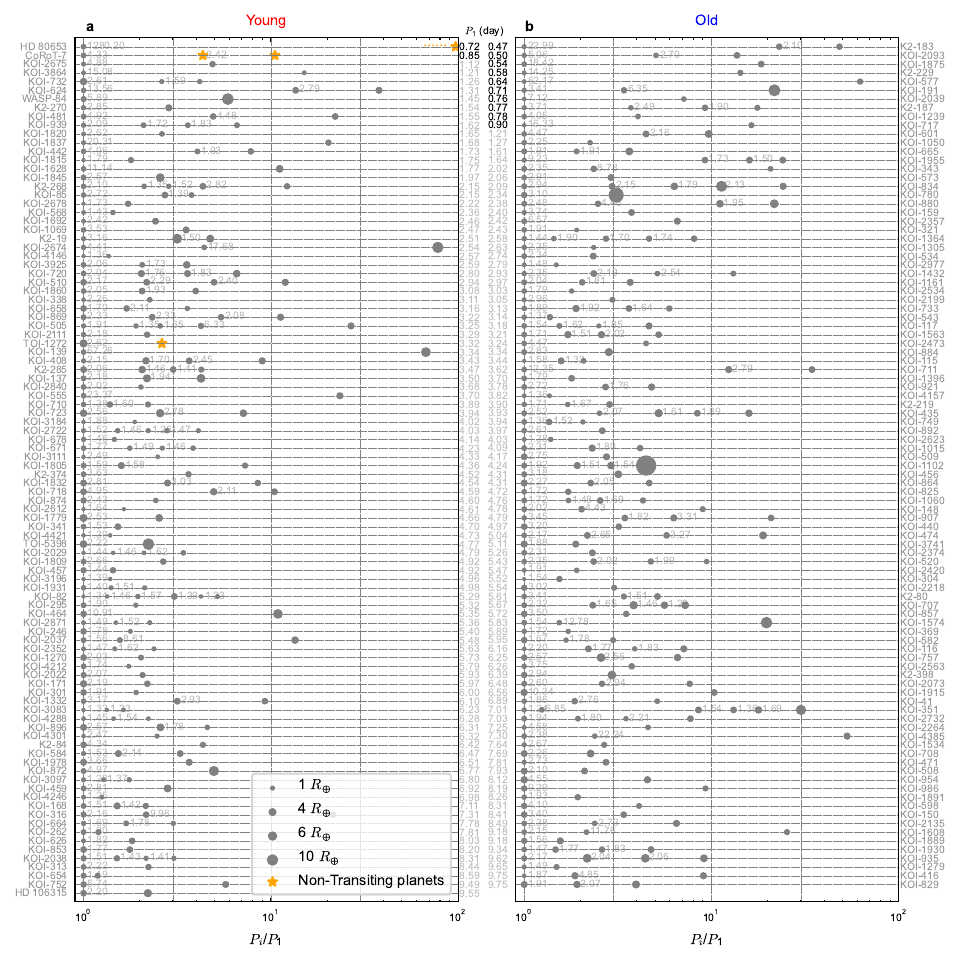}}
{\textbf{\hypertarget{figure_s15}{Supplementary Fig. 16: Similar to \hyperlink{figure_s15}{Supplementary Fig. 15}, but after applying the stellar parameter control to let the older subsample have comparable distributions in both $\rm [Fe/H]$ and $T_{\rm{eff}}$ as compared to the younger subsample.}}
Although the total sample size has reduced due to parameter control, the patterns observed before parameter control (\hyperlink{figure_s15}{Supplementary Fig. 15}) maintain, namely, \emph{(1) Planets are more spread out (i.e., larger orbital spacing) in older subsample (\textbf{b}) than in younger subsample (\textbf{a}), (see also in \hyperlink{6.2}{Methods \S6.2} and \hyperlink{figure_s18}{Supplementary Fig. 18}). (2) Systems with shorter innermost periods (i.e., $P_1<3-4$ days) tend to have a higher occurrence of non-transiting planets \citep[suggestive of larger mutual inclinations and consistent with observations by Dai et al. 2018,][]{2018ApJ...864L..38D} than systems with longer $P_1$. (3) Younger USP planet systems exhibit a larger fraction of non-transiting planets than older USP planet systems.}}
\end{figure*}

\clearpage

\begin{figure*}[!t]
\centerline{\includegraphics[width=\textwidth]{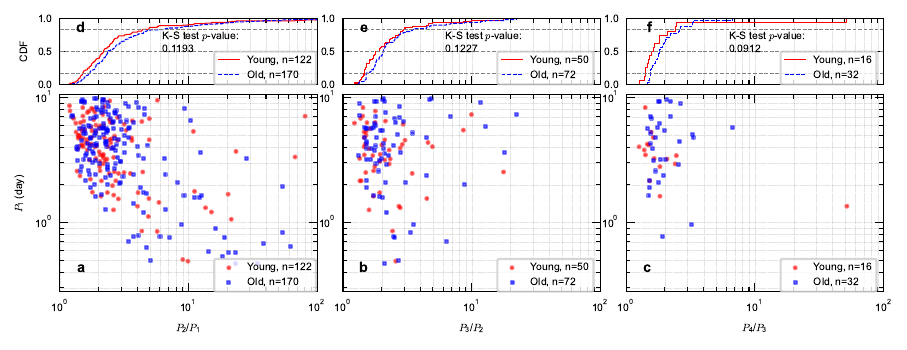}}
{\textbf{\hypertarget{figure_s17}{Supplementary Fig. 17: Comparison of the period ratio distributions of adjacent planetary pairs in the young (red) and old (blue) sub-samples before parameter control.}}
From left to right, inner pairs ($P_{2}/P_{1}$, \textbf{a and d}), middle pairs ($P_{3}/P_{2}$, \textbf{b and e}), and outer pairs ($P_{4}/P_{3}$, \textbf{c and f}).
The older subsample exhibits larger period ratios than the younger subsample in all three groups.
In each top panel \textbf{(d-f)}, we print the two sample K-S test $p-$values for the distributions of the older subsample compared to the younger subsample.}
\end{figure*}

\begin{figure*}[!t]
\centerline{\includegraphics[width=\textwidth]{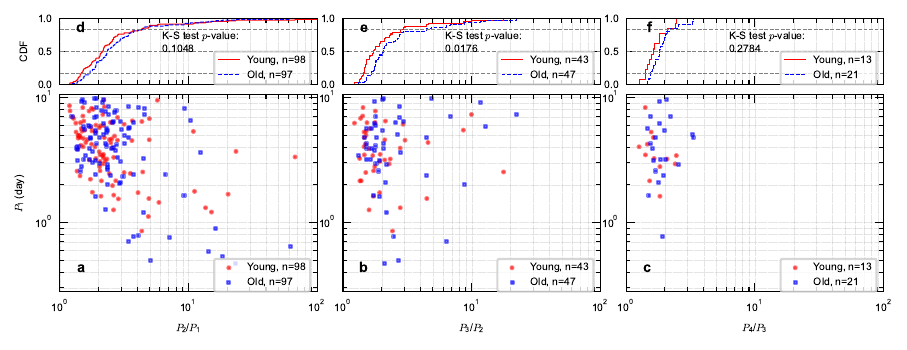}}
{\textbf{\hypertarget{figure_s18}{Supplementary Fig. 18: Similar to \hyperlink{figure_s17}{Supplementary Fig. 17}, but after applying the stellar parameter control to let the older subsample have comparable distributions in both $\rm [Fe/H]$ and $T_{\rm{eff}}$ as compared to the younger subsample.}}
After parameter control, the results that the older subsample exhibits larger period ratios than the younger subsample in all three groups remain.}
\end{figure*}

\clearpage

\hypertarget{Supplementary_discussion}{\subsection*{7. Supplementary discussion}}
\phantomsection\label{sec:Supplementary_discussion}
\hypertarget{7.1}{\subsubsection*{7.1 Validating the effects of detection completeness}}
\phantomsection\label{sec:7.1}

To eliminate the observational biases from different telescope sources and the effects of planet detection efficiency across different age bins, and to further clarify the age dependence of the occurrence and architecture of USP planetary systems, we use only {\it Kepler} sample to validate our results in the main text.

We retain only the {\it Kepler} sources from the hot planet host sample outlined in \hyperlink{Data_sample}{Methods \S1}.
We divide these {\it Kepler} planet systems into subsamples of different ages based on their hosts' relative probabilities between thick disk and thin disk ($TD/D$) and apply stellar parameter control to isolate the effect of age, as described in the \hyperlink{Frequency}{Methods \S5} and \hyperlink{architecture}{\S6}.
In parallel, we apply the same stellar selection criteria from \hyperlink{Data_sample}{Methods \S1} to the LAMOST-Gaia-{\it Kepler} kinematic value-added catalog \cite{2021AJ....162..100C} to select the {\it Kepler} parent stellar sample. 
We then divide the {\it Kepler} parent stellar sample into subsamples corresponding to the TD/D ranges mentioned above, aligning them with the division criteria applied to the {\it Kepler} planet systems and applying the same stellar parameter control.
We also compute the averaged detection efficiency curves for stellar subsamples of different ages using the \texttt{KeplerPORTs} code \cite[\url{https://github.com/nasa/KeplerPORTs,}][]{2015ApJ...809....8B,2017ksci.rept...19B} to validate the effects of detection completeness on subsequent results.

\hypertarget{7.1.1}{\subsubsection*{7.1.1 The frequency of Kepler USP planets as a function of age}}
\phantomsection\label{sec:7.1.1}

As shown in \hyperlink{figure_s19}{Supplementary Fig. 19a}, when the {\it Kepler} sample is divided into three subsamples: younger, intermediate, and older, the planet detection efficiencies around {\it Kepler} stars of different ages are comparable.
Thus, we infer that the detection efficiency has a minor impact on subsequent results. 
Subsequently, we recompute the relative frequency of USP planets among all hot small-mass planets ($f_{\rm USP}$) for different stellar ages (see \hyperlink{Frequency}{Methods \S5} for more details). 
According to the AVR, the kinematic ages of hosts in the {\it Kepler} subsamples are $0.41^{+0.04}_{-0.04}$ Gyr, $2.00^{+0.17}_{-0.14}$ Gyr, and $8.40^{+1.23}_{-0.96}$ Gyr for the younger, intermediate, and older subsamples, respectively. 
The corresponding $f_{\rm USP}$ values from the younger to the older subsample are $0.0676^{+0.0457}_{-0.0292}$ $(5/74)$, $0.0828^{+0.0299}_{-0.0227}$ $(13/157)$, and $0.1504^{+0.0417}_{-0.0333}$ $(20/133)$, respectively (\hyperlink{figure_s19}{Supplementary Fig. 19b}). 
The best fit for the kinematic sample is: 

\begin{equation}
f_{\rm USP} = 0.0623^{+0.0407}_{-0.0398} \times \log_{10}(t/\rm Gyr) + 0.0834^{+0.0222}_{-0.0227} \,
\tag{S1}
\label{Eq_M1}
\end{equation}

and $f_{\rm USP}$ is positively correlated with age in 9,407 out of 10,000 refitting sets, indicating a confidence level of 94.07\%.
The results still indicate that $f_{\rm USP}$ increases with age, with a slightly weaker confidence level possibly due to the reduced sample size, while the trend remains consistent with \hyperlink{figure_3}{Fig. 3} in the main text.

\begin{figure*}[!t]
\centerline{\includegraphics[width=\textwidth]{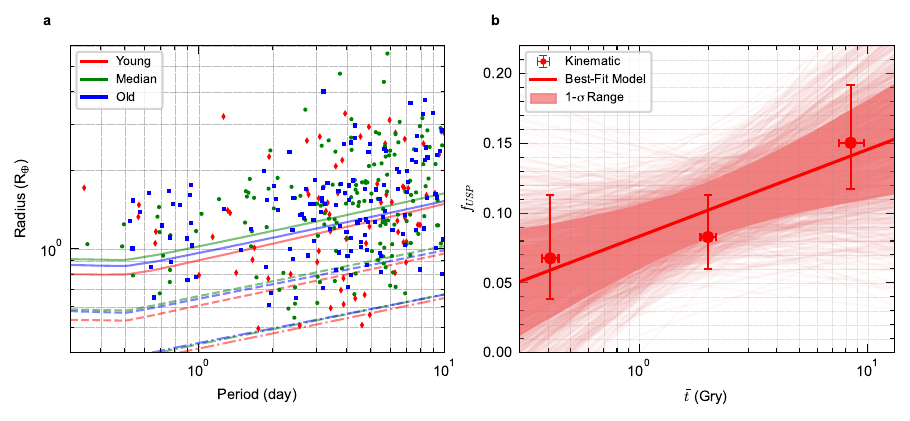}}
{\textbf{\hypertarget{figure_s19}{Supplementary Fig. 19: Orbital period and planetary radius for planets found in {\it Kepler} sample, and the frequencies of {\it Kepler} USP planets ($f_{\rm{USP}}$) as functions of kinematic ages.}}
\textbf{a}, Planets are divided into younger (red), intermediate (green), and older (blue) subsamples according to their hosts' age. 
The solid, dashed, and dashdot line represent the 90\%, 50\%, and 10\% average detection efficiency curves of these subsamples, respectively. 
The planet detection efficiencies around {\it Kepler} stars of different ages are comparable. 
\textbf{b}, The red solid lines and regions denote the best-fit linear model and 1 - $\sigma$ interval for the kinematic sample.
The thin line of light red represents the results of 10,000 fits from resampling the kinematic sample. 
The $f_{\rm{USP}}$ also increases with age in {\it Kepler} sample.
The best-fit linear model follows $f_{\rm USP} = 0.0623^{+0.0407}_{-0.0398} \times \log_{10}(t/\rm Gyr) + 0.0834^{+0.0222}_{-0.0227}$ (Equation \ref{Eq_M1} in Supplementary Discussion).}
\end{figure*}

\hypertarget{7.1.2}{\subsubsection*{7.1.2 Kepler USP planetary system architecture changes with time}}
\phantomsection\label{sec:7.1.2}

We also reinvestigate the age dependence of the architecture of {\it Kepler} USP planetary systems when the sample is divided into younger and older subsamples. 
As shown in \hyperlink{figure_s20}{Supplementary Fig. 20}, the detection efficiency curves of the younger and older subsamples are nearly indistinguishable. 
Therefore, the effects of detection efficiency can be regarded as negligible.
Furthermore, we reproduce the results of \hyperlink{figure_4}{Fig. 4} in the main text:

\begin{enumerate}

\item \emph{Period distribution.} The results remain an age-dependent period distribution (see \hyperlink{figure_s21}{Supplementary Fig. 21a}).
Although the K-S test result (0.3561) does not indicate a significant difference in the period distributions between the younger and older subsamples, the period distribution for the younger subsample still follows a power-law distribution, as demonstrated by a K-S test $p$-value of 0.9345. 
Conversely, the period distribution for the older subsample differs from a power-law distribution, as shown by a K-S test $p$-value of 0.1597.
Furthermore, a significant dip/pileup feature persists around $period=1$ day in the older subsample. 
the dip test conducted on the younger subsamples yields a $p-$value of 0.7900, corresponding to a confidence level of 79.00\% that supports the absence of a dip feature. 
On the other hand, for the older subsamples, the dip test reveals a $p-$value of 0.0478, which translates to a confidence level of $1-0.0478=95.22\%$ supporting the presence of a dip feature.

\item \emph{Orbital spacing.} The older subsample also exhibits a shift towards a larger period ratio (with a median of 2.16) compared to the younger subsample (with a median of 1.90). 
The KS test of these two period ratio distributions yields a $p-$value of 0.0106, indicating that the two distributions are distinct at a confidence level of $1-0.0106=98.94\%$ (see \hyperlink{figure_s21}{Supplementary Fig. 21b}). 

\item \emph{Transiting planet multiplicity.} As shown in \hyperlink{figure_s21}{Supplementary Fig. 21c}, the younger {\it Kepler} USP planet systems contain a lower fraction of multiple transiting systems. 
For HS planet systems, the fraction of multiple transiting systems remains comparable between the younger ($86/159\sim54.09^{+5.83}_{-5.83}\%$) and older ($81/150\sim54.00^{+6.00}_{-6.00}\%$) subsamples. 
Conversely, for USP planet systems, the younger subsample ($0/11, \sim0.00^{+16.73}\%$) has a much lower fraction of multiple transiting systems than the older subsample ($6/20, \sim30.00^{+17.91}_{-11.90}\%$). 

\end{enumerate}

Overall, we still observe the age-dependent variations in the architecture of USP planetary systems in the {\it Kepler} sample. 
Although the confidence level shows varying degrees of decline, possibly due to the reduced sample size, the results are overall consistent with those presented in \hyperlink{figure_4}{Fig. 4} of the main text.

\hypertarget{7.1.3}{\subsubsection*{7.1.3 Section summary}}
\phantomsection\label{sec:7.1.3}

In this section, we validate the effects of detection completeness on our results using a more homogeneous {\it Kepler} sample. 
The validation indicates that the planet detection efficiencies around stars of different ages in our {\it Kepler} sample are comparable.
Meanwhile, although the sample size is reduced, the results from the {\it Kepler} sample remain relatively consistent with those from the multi-source sample.
Our validation demonstrates that the observed age dependence of the occurrence and architecture of USP planetary systems is not caused by observational biases but is instead driven by age.
Further studies and detections will help clarify this issue and deepen our understanding of the formation and evolution of USP planet systems.

\begin{figure*}[!t]
\centerline{\includegraphics[width=0.7\textwidth]{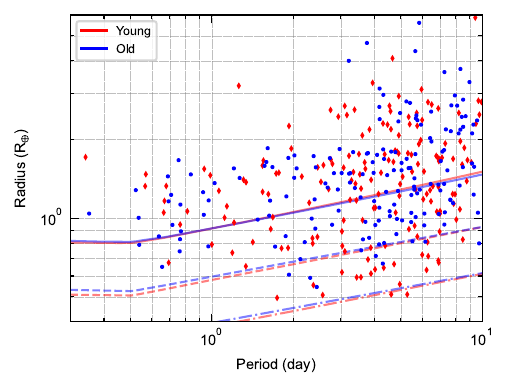}}
{\textbf{\hypertarget{figure_s20}{Supplementary Fig. 20: Orbital period and planetary radius for planets found in {\it Kepler} sample.}} 
Planets are divided into younger (red), and older (blue) subsamples according to their hosts' age.
The solid, dashed, and dashdot line represent the 90\%, 50\%, and 10\% average detection efficiency curves of these subsamples, respectively.
The detection efficiency curves of the younger and older subsamples are nearly indistinguishable.}
\end{figure*}

\begin{figure*}[!t]
\centerline{\includegraphics[width=\textwidth]{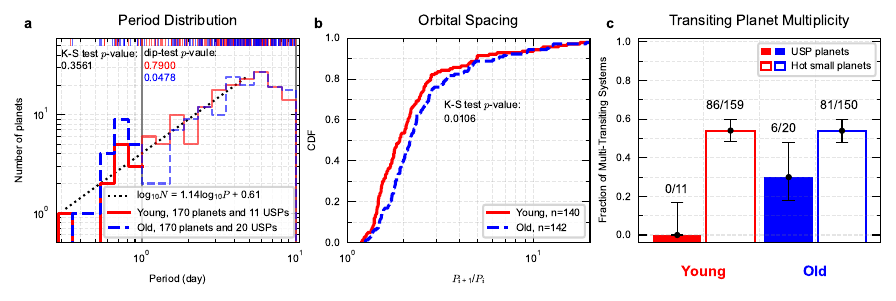}}
{\textbf{\hypertarget{figure_s21}{Supplementary Fig. 21: Similar to \hyperlink{figure_4}{Fig. 4} in the main text, but using {\it Kepler} sample.}}}
\end{figure*}

\end{document}